\documentclass[pdftex,twocolumn,epjc3]{svjour3}          

\RequirePackage[T1]{fontenc}

\smartqed  
\usepackage{amssymb}
\RequirePackage{graphicx}
\RequirePackage{mathptmx}      
\RequirePackage{amsmath}
\RequirePackage{flushend}
\RequirePackage[numbers,sort&compress]{natbib}
\RequirePackage[colorlinks,citecolor=blue,urlcolor=blue,linkcolor=blue]{hyperref}

\journalname{Eur. Phys. J. C}

\begin{document}

\title{Melting of scalar mesons and black-hole quasinormal modes in a holographic QCD model}

\author{Luis A. H. Mamani\thanksref{e1,addr1,addr2}, 
        Alex S. Miranda \thanksref{e2,addr2}
        \and
        Vilson T. Zanchin\thanksref{e3,addr1} 
}

\thankstext{e1}{e-mail: luis.mamani@ufabc.edu.br}
\thankstext{e2}{e-mail: asmiranda@uesc.br}
\thankstext{e3}{e-mail: zanchin@ufabc.edu.br}

\institute{
Centro de Ci\^encias Naturais e Humanas,
Universidade Federal do ABC,\\
Rua Santa Ad\'elia 166, 09210-170, Santo Andr\'e, S\~ao Paulo, Brazil\label{addr1}
          \and
Laborat\'orio de Astrof\'{\i}sica Te\'orica e Observacional\\ 
Departamento de Ci\^encias Exatas e Tecnol\'ogicas\\
Universidade Estadual de Santa Cruz, 45650-000, Ilh\'eus, Bahia, Brazil\label{addr2}
}

\date{Received: date / Accepted: date}

\maketitle

\begin{abstract}
A holographic model for QCD is employed to investigate the effects of the
gluon condensate on the spectrum and melting of scalar mesons.
We find the evolution of the free energy density with the temperature, and the result shows that
the temperature of the confinement/de\-con\-fine\-ment transition is sensitive to the
gluon-condensate parameter.
The spectral functions (SPFs) are also obtained and show a series of peaks in the low-temperature regime,
indicating the presence of quasiparticle states associated to the mesons, while the number of
peaks decreases with the increment of the temperature, characterizing the quasiparticle melting.
In the dual gravitational description, the scalar mesons are identified with the black-hole quasinormal
modes (QNMs). We obtain the spectrum of QNMs and the dispersion relations corresponding to the scalar-field
perturbations of the gravitational background, and find their dependence with the gluon-condensate parameter.
\end{abstract}

\section{Introduction}
It is well known in quantum chromodynamics (QCD) that the gluon condensate
has a relevant role in the low-energy dynamics; in the absence of
quarks it represents the vacuum of QCD \cite{Shuryak:1988ck}. 
It is also known that the low-energy regime of QCD cannot
be studied with the usual mathematical tools employed in 
the regime of high energies, where the coupling constant is small and one
can use perturbative techniques to obtain information about the system.
Differently from the high-energy regime, the coupling constant is large
for low energies and perturbative techniques are no longer applicable. 
There are some approaches attempting to obtain information about the low-energy dynamics.
One of these approaches is the lattice field theory, where the problems are discretized
and solved on a lattice, using powerful computational resources (see e.g. Ref.~\cite{Lucini:2013qja}
for a review on this subject). Another technique is the operator product expansion (OPE),
also called the SVZ (Shifman, Vainshtein, and Zakharov) sum rules, 
which tries to extend some perturbative results to the low-energy regime \cite{Shifman:1978bx}.
Both approaches have limitations and cannot provide a full real-time description of the
low-energy regime of QCD.

In recent years, as an alternative tool,
a variety of gravitational holographic models
has been used
to study the non-perturbative regime of QCD.
In this context, there are two main roads to follow. The so-called top-down approach has a ten- or eleven-dimensional
superstring solution as starting point. After some compactifications, it is obtained a five-dimensional effective
gravitational model, which is dual to a four-dimensional conformal field theory (CFT) living at the boundary of a
curved spacetime with a negative cosmological constant.
This curved spacetime is known as anti-de Sitter or AdS for short (for a review on top-down models, see for instance
Ref.~\cite{Kim:2012ey}). In the second approach, the so-called bottom-up models are designed so that the dual gravitational
theory reproduces some known results or properties in QCD.
In these models, the gravitational dual theory does not need to follow as a classical limit of a
superstring theory (for a review on bottom-up models, see Ref.~\cite{Erlich:2014yha}).

The main advantage of the gauge/gravity duality, formulated in its original form in \cite{Maldacena:1997re,Gubser:1998bc,Witten:1998qj}, 
is that we can map problems in a strongly coupled field theory, living in a flat $d$-dimensional spacetime, to problems in a
$(d+1)$-dimensional classical theory of gravity. This map is implemented by associating each local operator in the quantum field theory
to a classical field in the gravitational side, i.e., $\mathcal{O}(x^{0},\cdots,x^{d}) \leftrightarrow X(x^{0},\cdots,x^{d},x^{d+1})$. 
For example, in four-dimensional QCD the operator Tr$\,F^2$, that characterizes the scalar sector of the gluon field $F_{\mu\nu}$, is
dual to a scalar field $\Phi$, known as the dilaton. This correspondence is such that the value of the dilaton at the boundary,
$\Phi(x^{0},\cdots,x^{4},x^{5}=z=0)=\phi_0(x^{0},\cdots,x^{4})$, is a source for the operator Tr$\,F^2$, which means
they are coupled as $\int dx^{4}\,\phi_0(x) \text{Tr}\,F^2(x)$.

The first attempt to take into account the gluon condensate in a holographic bottom-up approach for QCD was developed by 
Csaki and Reece \cite{Csaki:2006ji}. By considering a dilaton that couples to the operator Tr$\,F^2$ at the AdS boundary,
they showed that the  asymptotic behaviour of this field close to the boundary must be of the form $\Phi(z)=\phi_0+G\,z^4$,
where $z$ is the holographic coordinate. This result is consistent with the general asymptotic solution for a massless scalar
field close to the boundary: 
$\Phi(z)=\phi_0\,z^{4-\Delta}+G\,z^\Delta$, where $\phi_0$, $G$ and $\Delta=4$ are interpreted,
respectively, as the source, the vacuum expectation value (VEV) and the conformal dimension of the operator Tr$\,F^2$
(the VEV is also known as the gluon condensate). In spite of a relative success with the introduction of the gluon condensate,
the authors of Ref.~\cite{Csaki:2006ji} obtained a scalar glueball spectrum which does not follow a linear Regge trajectory.

Nevertheless, there are some well-succeeded approaches to QCD, such as the soft-wall model \cite{Karch:2006pv} and the 
improved holographic model \cite{Gursoy:2007cb,Gursoy:2007er}, which consider a quadratic dilaton in the IR and
thus guarantee a linear behaviour for the glueball and meson spectra. Motivated by these works, in this paper
we implement a model
with a dilaton field which is quartic in the UV (to describe correctly the gluon condensate) and quadratic in the IR (to guarantee
linear  behavior of the spectrum). As in the soft-wall model, the background metric is fixed, and the conformal
symmetry is explicitly broken with the introduction of an exponential dilaton dependent term in the five-dimensional action. 
Such a term does not modify the gravitational background and generates a dual field-theory
energy-momentum tensor with nonvanishing trace.

Among the important quantities that are obtained using our model are the
spectral functions (SPFs), which are fundamental for the understanding of
the hadronic properties and the vacuum structure of QCD. Additionally, the SPFs at finite temperature
may shed some light
about the influence of the surrounding medium on the hadronic internal structure. In Refs.~\cite{Asakawa:2000tr,Asakawa:2002xj} the
maximum entropy method was used to construct the SPFs in lattice QCD. In holography, a dual description of a
finite temperature field theory is obtained by considering a black hole in the gravitational AdS background \cite{Witten:1998zw}.
The authors of Ref.~\cite{Son:2002sd} developed a prescription to calculate retarded Green functions from the 
five-dimensional on-shell action, allowing to get information of the finite-temperature field theory
from quantities defined on the gravitational side. As an example we mention the hydrodynamic transport properties
obtained in Refs.~\cite{Policastro:2002se,Policastro:2002tn}, where the poles of the low-momentum
limit of retarded Green functions are identified with the low-lying quasinormal modes (QNMs).
In fact, it was shown that the poles of the
finite-temperature correlation functions are related to the full spectrum of QNMs, not just in the hydrodynamic regime. 
From the gravitational point of view, the spectrum of QNMs are produced by perturbations on the AdS black-hole
spacetime that satisfy specific boundary conditions: an ingoing wave condition at the horizon
and a Dirichlet condition at the boundary of the AdS spacetime.
Interestingly, in recent years it has been an increasing interest on the nonhydrodynamic QNMs,
because these modes dominate the dynamics of fluctuations before the system reaches the local hydrodynamic equilibrium
(see Refs.~\cite{Noronha:2011fi,Heller:2014wfa,Janik:2014zla} for a discussion). 
In addition, the study of the nonhydrodynamic 
QNMs may shed some light on the understanding of early time dynamics of 
some strongly coupled field theory models (for example, the QCD quark-gluon plasma).
For discussion and details on quasinormal modes, see  Refs.~\cite{Berti:2009kk,Konoplya:2011qq}.

The present paper is organized as follows.
In Sec.~\ref{Sec:ScalarMesons} we introduce the action and the equations of motion of the model
and define the dilaton field that interpolates between the UV and IR. 
Then, we obtain the mass spectrum of mesons at zero temperature and
explore the effects of the parameter associated with the gluon condensate on this spectrum. 
Sec.~\ref{Sec:ScalarMesonsFiniteT} is devoted to the study
of the finite-temperature effects and, to do that,
we calculate the free energy density of the thermal and black-hole
states. Yet in Sec.~\ref{Sec:ScalarMesonsFiniteT} we 
obtain the SPFs and make explicit the 
dependence of the results on the gluon condensate and 
temperature. 
In Sec.~\ref{Sec:QNMs} we present the 
spectrum of QNMs calculated by using three different numerical 
techniques: power series, Breit-Wigner, and 
pseudo-spectral methods. A discussion and comparison between the methods are also presented.
To complement the numerical analysis, in 
Sec.~\ref{Sec:DispersionRelations} we present and discuss 
the results 
of the dispersion relations. 
In Sec.~\ref{Sec:conclusion} we conclude with some remarks and the final comments.

\section{Scalar mesons from holographic QCD}
\label{Sec:ScalarMesons}

In this section we explore a holographic description
of the scalar sector of the mesons. This is done by employing the action proposed in 
Ref.~\cite{Karch:2006pv}, whose respective approach is known as the 
soft-wall model. It is a five-dimensional effective action
with two scalar fields, the dilaton $\Phi(z)$ and a
scalar $S(x^{\mu},z)$, both introduced as probe fields, 
which means that the backreaction of the fields on the geometry
is neglected.  The holographic dictionary \cite{Maldacena:1997re} 
establishes that the dilaton is dual to the operator Tr$\,F^2$ 
\cite{Csaki:2006ji}, while the scalar
field $S$ is dual to the operator $\bar{q}q$ 
\cite{DaRold:2005vr,Gherghetta:2009ac}.

\subsection{Equations of motion for the fields}
\label{SubSec:EqOfMotion}
We start with the metric that describes the five-dimensional anti-de Sitter spacetime.
Such a metric is a solution of the Einstein equation with a negative cosmological
constant, and takes the following form in Poincar\'e coordinates:
\begin{equation}\label{ThermalAdSMetric}
d\tilde{s}^{\,2}
=\tilde g_{mn}dx^{m}dx^{n}
=e^{2A(z)}\left(dz^2+\eta_{\mu\nu}dx^{\mu}dx^{\nu}\right),
\end{equation}
where the five-dimensional spacetime metric has the signature $(-,+,+,+,+)$.
The warp factor is defined by $A(z)=\log{\left(\ell/z\right)}$, where
$\ell$ is the AdS radius. From now
on, in this work, we set the radius $\ell=1$ to simplify the notation. 
In the system of coordinates $\{x^{\mu},\,z\}$, the boundary field theory lies at
$z=0$, which is identified as the UV fixed point, while 
$z\to \infty$ is the deep IR region. This identification
is possible because the warp factor and the energy scale $E$ of the dual field 
theory are related by $e^{A(z)}=E$ \cite{Gursoy:2007cb,Gursoy:2007er}.

The five-dimensional action, that describes chiral symmetry 
breaking in the meson sector and contains $SU(2)_{L}\times SU(2)_{R}$ 
gauge fields with a bifundamental scalar field $X$, can be written as
\cite{Karch:2006pv} 
\begin{equation}\label{OriginalAction}
\begin{split}
\mathcal{S}_{5D}=&-\int d^5x\sqrt{-\tilde{g}}\, e^{-\Phi}\,
\text{Tr}\bigg[|DX|^2+m_{X}^2 |X|^2\\
&+\frac{1}{4g_5^2}\left(F^2_L+F^2_R\right)\bigg],
\end{split}
\end{equation}
where $D^mX=\partial^mX-iA_{L}^{m}X+iXA_{R}^{m}$ and
$F_{L,R}^{mn}=\partial^{m}A^{n}_{L,R}-\partial^{n}A^{m}_{L,R}-i\left[A^{m}_{L,R},A^{n}_{L,R}\right]$,
with $A_L$ and $A_R$ the gauge fields and $X$ the scalar field (tachyon) responsible for the
chiral symmetry breaking $SU(2)_{L}\times SU(2)_{R}\to SU(2)_{V}$ (for details, see Ref.~\cite{Erlich:2005qh}).
Following Refs.\cite{DaRold:2005vr,Gherghetta:2009ac} the 
scalar sector of the foregoing action can be obtained by turning 
off all the background fields except the fluctuation of the 
bifundamental field, which is decomposed in the 
form $X=X_0(z)+S(x,z)$, with $S(x,z)$ being a perturbation on the 
background value $X_0(z)$.
Hence, the action that describes the scalar sector
of the mesons is given by
\begin{equation}\label{ScalarAction}
\mathcal{S}_{5D}=-\int dx^5\, 
\sqrt{-\tilde{g}}\,e^{-\Phi}
\left[\left(\partial_{m}S\right)\left(\partial^{m}S\right)
+m_{X}^2\,S^{2}\right],
\end{equation}
where $m_{X}^2=-3$ is the mass of the scalar field. 
The dimension of the operator dual to the scalar field 
$S(x,z)$ satisfies the equation $\Delta(\Delta-4)=m_{X}^2$ \cite{Gubser:1998bc}.
Solving this algebraic equation, we get
$\Delta=3$, which means the operator dual to $S$
has dimension $3$. 

The equation of motion obtained from varying the action 
\eqref{ScalarAction} can be recast into a Schr\"odinger-like 
form by introducing the Fourier decomposition 
$S(x,z)=S(z)e^{-ik_\mu x^{\mu}}$, and a Bogoliubov transformation 
$S(z)=e^{-B(z)}\psi(z)$, where the auxiliary function 
$2B(z)=3A(z)-\Phi(z)$ was introduced. In doing so, we get
\begin{equation}\label{SchroEq}
-\psi''(z)+V(z)\psi(z)=m_s^{2}\psi(z),
\end{equation}
where $m^{2}_{s}=-k_{\mu}k^{\mu}$ and $'$ indicates $d/dz$.
In terms of the dilaton field, the effective potential takes the form
\begin{equation}\label{EffPotential}
V(z)=\frac{3}{4z^2}+\left[\frac{3}{2z}
+\frac{\Phi'(z)}{4}\right]\Phi'(z)-\frac{\Phi''(z)}{2}.
\end{equation}
We now turn attention to the dilaton field $\Phi(z)$ and
consider a functional dependency on $z$ that is consistent with some
aspects of QCD, namely, $\Phi(z) \sim z^4$ in the UV (to describe correctly
the gluon condensate) \cite{Csaki:2006ji}, and $\Phi(z) \sim z^2$ in 
the IR (to guarantee confinement) 
\cite{Gursoy:2007er}. To smoothly connect these regimes,
we use an interpolation function in the form \cite{Ballon-Bayona:2017sxa}
\noindent
\begin{equation}\label{DilaEq}
\Phi(z)=\phi_{0}+\frac{G\, z^{4}}{1+\frac{G}{C}\, z^{2}},
\end{equation}
such that the asymptotic behavior of $\Phi(z)$ close to $z=0$ becomes
\noindent
\begin{equation}\label{DilatonUVEq}
\Phi(z)=\phi_{0}+G\, z^{4}+\cdots,\qquad z\to 0,
\end{equation}
\noindent
where $\phi_{0}$ is the source that couples to the 
dual operator Tr$\,F^{2}$, $G$ is the energy scale associated with 
the gluon condensate, and
the ellipses indicate subleading terms. 
In the deep IR region, the dilaton 
field goes like
\noindent
\begin{equation}\label{DilatonIREq}
\Phi(z)=C\, z^{2}+\cdots \qquad z\to \infty,
\end{equation}
\noindent
where $C$ characterizes the confinement energy scale and
the ellipses mean subleading contributions. The parameter 
$C$ can be chosen by matching the smaller eigenvalue of the 
vectorial sector of the action \eqref{OriginalAction} with the experimental mass of the
lightest $\rho$ meson. This was done in Ref. \cite{Herzog:2006ra}, and the
resulting value is $C=0.151(GeV)^2$. In the 
forthcoming sections we are not going to fix the parameters, 
since the aim of this paper is to show qualitative instead 
of quantitative results associated to this holographic model.

For numerical purposes, we rewrite Eq.~\eqref{DilaEq}
in terms of the dimensionless coordinate $u=z\sqrt{C}$, 
\noindent
\begin{equation}\label{NewDilaEq}
\Phi(u)=\phi_0+\frac{\mathcal{G}u^{4}}{1+\mathcal{G}u^{2}},
\end{equation}
\noindent
where the information of the gluon condensate is now 
contained in the dimensionless parameter
$\mathcal{G}=G/C^{2}$. After the change of coordinate 
$z\rightarrow u$, the effective potential
and the mass $m_{s}$ in the Schr\"odinger-like  
equation are normalized
by the parameter $C$ as $\hat{V}=V/C$ and 
$\hat{m}_s=m_s/\sqrt{C}$. In the next sections,
we are going to investigate if (and how) the spectrum of 
scalar mesons depends on the parameter $\mathcal{G}$.

\subsection{Analysis of the zero-temperature effective potential}

Here we point out the differences
between our approach and the original soft-wall model \cite{Karch:2006pv}.
As commented above, we are considering a dilaton field which is quartic  
in the UV and quadratic in the IR. This difference in relation to 
the soft-wall model, 
where the dilaton is quadratic from the UV to IR, requires the 
introduction of a new free parameter related to the energy scale that characterizes the gluon condensate \cite{Csaki:2006ji}.
The dimensionless version of this new parameter is 
identified with $\mathcal{G}$ in Eq. \eqref{NewDilaEq}.

The asymptotic behaviour of the potential in the UV depends 
on  $\mathcal{G}$.
This statement can be appreciated expanding the potential 
(\ref{EffPotential}) near the boundary,
\noindent
\begin{equation}\label{UVPotentialEq}
V(u)=\frac{3}{4u^{2}}+6\,\mathcal{G}^{2}u^{4}
+4\,\mathcal{G}^{2}\left(1-4\,\mathcal{G}\right)u^{6}+\cdots
\qquad u\to 0,
\end{equation}
\noindent
where the ellipses indicate subleading terms, 
expressed as higher order powers of $u$.

On the other hand, the asymptotic form of the 
potential (\ref{EffPotential})
in the deep IR region can be written as
\noindent
\begin{equation}\label{IRPotentialEq}
V(u)=u^{2}+\cdots\qquad u\to \infty,
\end{equation}
\noindent
where the ellipses represent subleading terms, 
suppressed as powers of $1/u$.

\begin{figure}[!ht]
\centering
\includegraphics[width=8.2cm,angle=0]{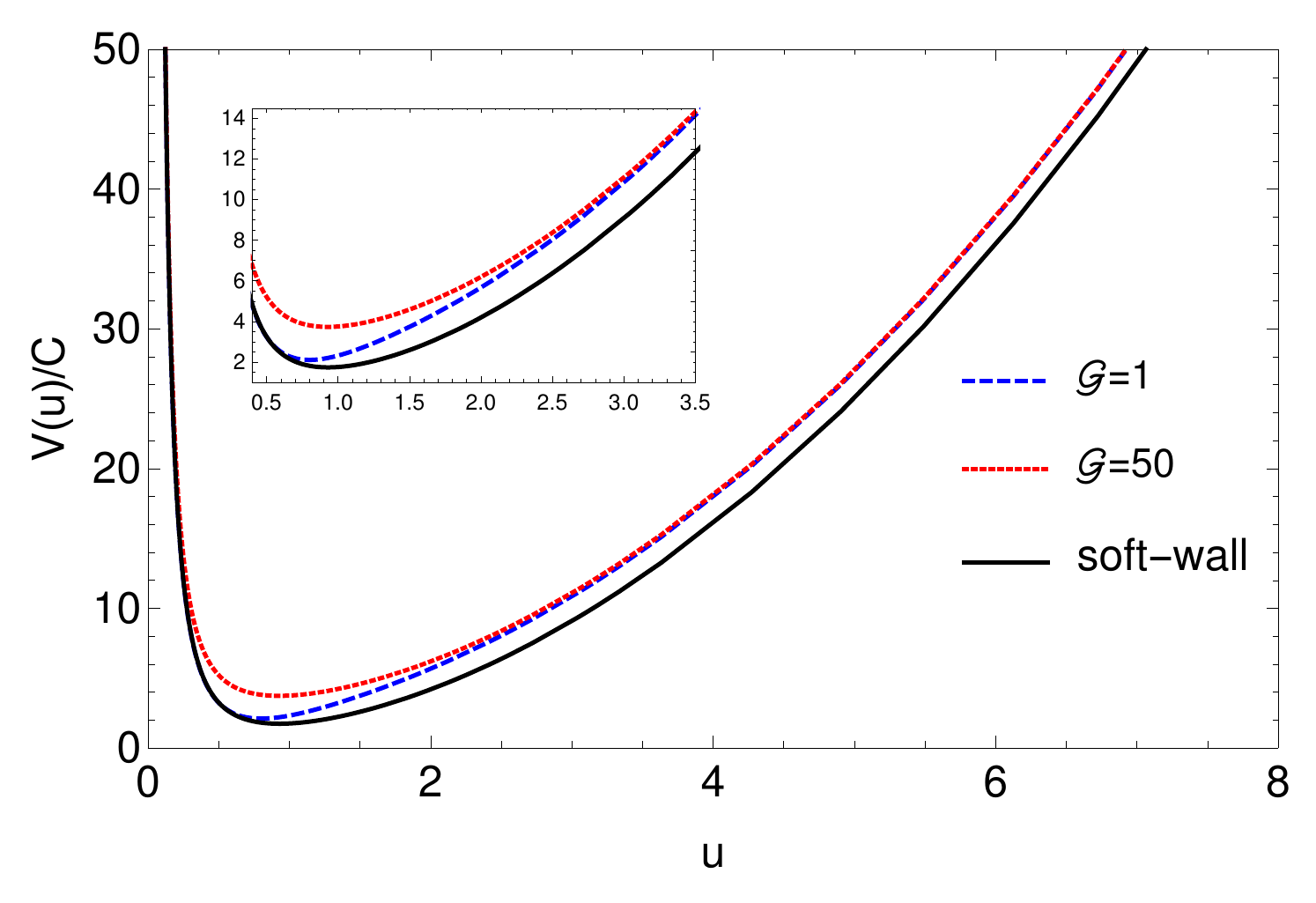} 
\caption{
The curves of the potential of Eq.~(\ref{EffPotential})
for two different values of $\mathcal{G}$ and, for comparison, the curve of the original potential in the soft-wall 
model \cite{Karch:2006pv}. The asymptotic 
behaviours are similar and the main difference between the 
models lies in the intermediate region.}
      \label{Potential}
\end{figure}

By comparing the above results with those of the original 
soft-wall model \cite{Karch:2006pv},
we notice the influence of the quartic dilaton 
on the asymptotic UV behavior of the effective potential,
which depends on the value of the gluon-condensate dimensionless parameter $\mathcal{G}$.
To see  this difference quantitatively, we plot in  
Fig.~\ref{Potential} the potential obtained in the original 
soft-wall model \cite{Karch:2006pv}
and the numerical results obtained from Eq.~(\ref{EffPotential}).
\noindent

As expected, the asymptotic behaviors are similar and the 
main differences between our approach and 
the original soft-wall model
lies in an intermediate region between
the UV and IR, as it can be seen in Fig.~\ref{Potential}.

\subsection{Asymptotic solutions for the scalar field}

Once the asymptotic behaviour of the effective potential is known,
we can now find the asymptotic solutions of the 
Schr\"odinger-like equation~\eqref{SchroEq}.
The two regimes, UV and IR, are dealt with separately. 

To obtain the asymptotic solutions in the UV regime,
we first replace the leading terms of the potential
from Eq.~\eqref{UVPotentialEq} into Eq.~\eqref{SchroEq}, 
which leads to 
\noindent
\begin{equation}
-\psi''(u)+
\left(\frac{3}{4u^{2}}+6\,\mathcal{G}^{2}u^{4}\right)\psi(u)
=\hat{m}_s^{2}\psi(u),\qquad u\to 0. 
\end{equation}
\noindent
As this is a second-order differential equation we have 
two solutions close the boundary
\noindent
\begin{equation}\label{AsymptSol}
\psi(u)=c_{1} u^{-1/2}+c_2 u^{3/2},\qquad u\to 0.
\end{equation}
\noindent
The first term on the right hand side of \eqref{AsymptSol} 
is the non-normalizable solution (see Appendix of
Ref.~\cite{Ballon-Bayona:2017sxa} for details), while the second term
is the normalizable one. As we are uniquely interested in 
normalizable solutions, we set $c_1=0$.

One might be surprised about the fact that the 
parameter $\mathcal{G}$ does not affect the asymptotic solutions 
\eqref{AsymptSol}. This is because the 
asymptotic solution of the scalar field $S(z)$
is not affected by a quartic 
dilaton in the UV. In fact, its contribution is 
subleading in Eq.~\eqref{AsymptSol}.

In the deep IR regime the
asymptotic behaviour of the potential is given by 
Eq.~\eqref{IRPotentialEq} and
the Schr\"odinger-like equation~\eqref{SchroEq} reduces to
\noindent
\begin{equation}
-\psi''(u)+u^{2}\psi(u)=\hat{m}_s^{2}\psi(u),\qquad u\to \infty. 
\end{equation}
\noindent
Hence, the solution that guarantees convergence of the 
wave function in this region is given by 
\noindent
\begin{equation}\label{AsymptSolIR}
\psi(u)=b_1 u^{(\hat{m}_{s}^{2}-1)/2} e^{-u^{2}/2},
\qquad u\to \infty.
\end{equation}
\noindent
This function, which gives the asymptotic behaviour of the wave function $\psi(u)$
in the IR region, and the solution \eqref{AsymptSol}, that represents the asymptotic
behaviour of $\psi(u)$ in the UV region, are used below in the search for a full solution
of Eq.~\eqref{SchroEq} through numerical methods.

\subsection{Analysis of the mass spectrum}
\label{SubSec:Spectrum}

As pointed out above, the normalizable solutions of the differential 
equation~\eqref{SchroEq} are associated to scalar-meson states. We 
obtain these solutions by solving
numerically the Schr\"odinger-like equation~\eqref{SchroEq} with a shooting method.
For the numerical integration, boundary conditions need to be provided.
In the present case, we use the near-boundary UV ($u=0$) normalizable solution
\eqref{AsymptSol}
and its derivative as ``initial" conditions, and complement them by imposing regularity of the wave function in the deep IR region. 
These conditions are satisfied just for a discrete set of values of the mass parameter $\hat m_{s}^{2}$.
Notice that, alternatively, we might use the asymptotic solution \eqref{AsymptSolIR} in the IR ($u\to\infty$)
and its derivative as ``initial" conditions, and require a regular behaviour of the wave function in the UV boundary $u=0$. 
Both approaches give the same solutions to the eigenvalue
problem. 

\begin{figure}[!ht]
\centering
\includegraphics[width=8.2cm,angle=0]{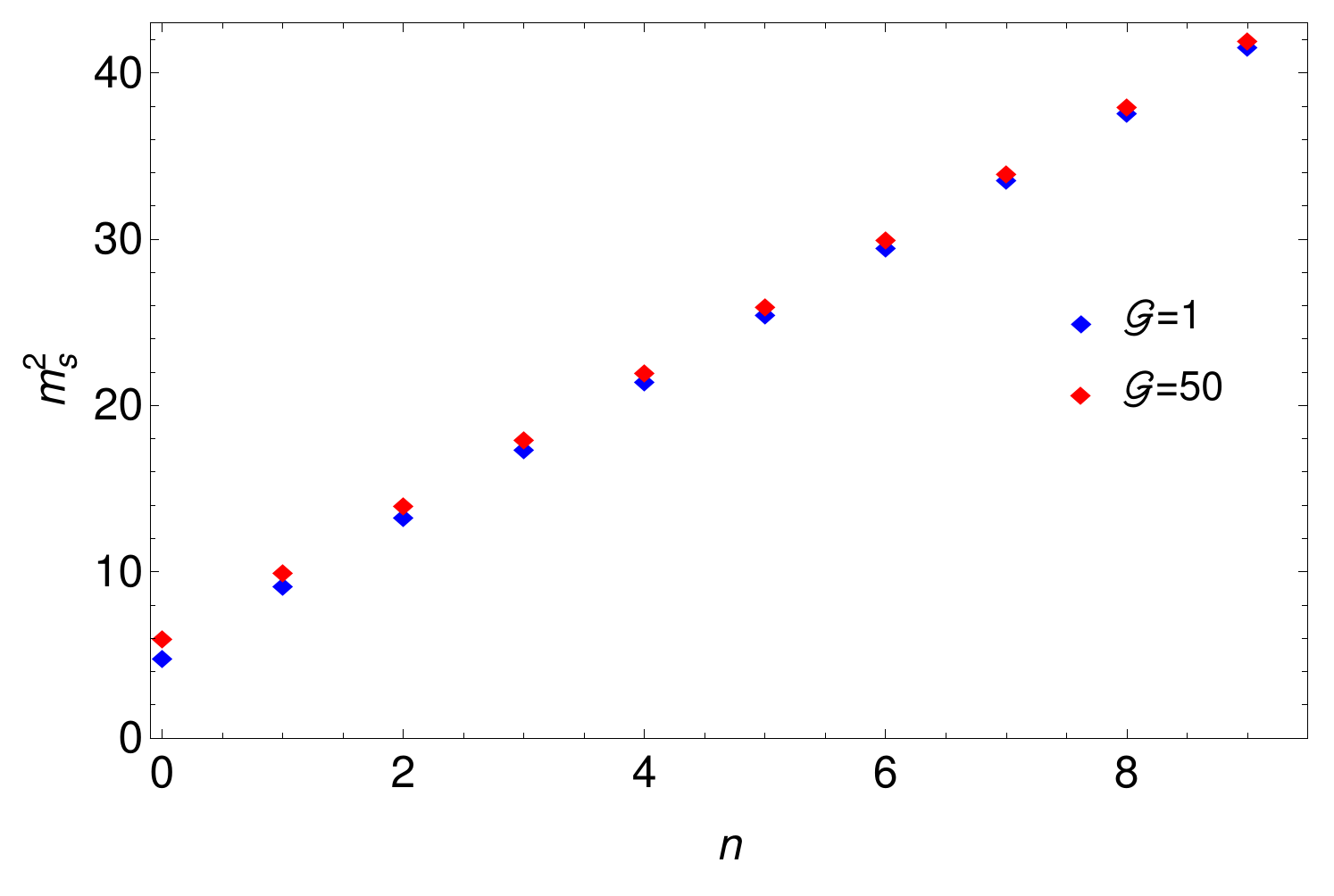} 
\caption{
The mass spectrum of scalar mesons at zero 
temperature for two different values of $\mathcal{G}$.}
\label{ScalarSpectrumT0G1G50}
\end{figure}

In Table \ref{Tab:01} we present the
first nine eigenvalues $\hat m_{s}^{2}$ for two different values of the 
parameter $\mathcal{G}$; namely, $\mathcal{G}=1$ and $\mathcal{G}=50$.
These particular values are chosen arbitrarily and their separation is taken large
enough to clearly display the dependence of the mass spectrum on 
such a parameter.
\begin{table*}[ht!]
\begin{center}
\begin{tabular}{l|c|c|c|c|c|c|c|c|c|c}
\hline 
\hline
 $n$ & 0 & 1 & 2 & 3 & 4 & 5 & 6 & 7 & 8 & 9  \\ \hline 
 $\hat m_{s}^{2}(\mathcal{G}=1)$ & 4.84 & 9.16 & 13.31 & 17.40 & 21.46
 & 25.51 & 29.55 & 33.57 & 37.60 & 41.62
 \\ \hline 
 $\hat m_{s}^{2}(\mathcal{G}=50)$ & 5.99 & 9.99 & 13.99 & 17.99 & 
 21.99 & 25.99 & 29.99 & 33.99 & 37.99 & 41.99  \\
 \hline\hline
\end{tabular}
\caption{The mass spectrum of scalar mesons at zero 
temperature for two different values of the dimensionless parameter 
$\mathcal{G}$. The masses are dimensionless.} 
\label{Tab:01}
\end{center}
\end{table*}
These results show that the lower excited states are more sensitive to the change of the
parameter $\mathcal{G}$ than the higher excited states. 
To make this fact transparent, we plot the data of 
Table \ref{Tab:01} in Fig.~\ref{ScalarSpectrumT0G1G50}, where the 
differences among the two spectra at low masses is clearly seen.
For comparison, let us mention the previous results found in the 
original soft-wall model
\cite{Karch:2006pv,Vega:2008af, Colangelo:2008us}, 
whose mass spectrum has the closed form
\begin{equation}\label{MassLiterature}
\hat m_s^2=(4\,n+6), \qquad n=0,1,2,\cdots\,.
\end{equation}
\noindent
Numerical fits to our results yield the following 
expressions:
\noindent
\begin{equation}
\begin{split}
\hat m_s^2(\mathcal{G}=1)=&4.05\,n+5.17,\qquad\;n=0,1,2,\cdots\\
\hat m_s^2(\mathcal{G}=50)=&4.00\,n+5.99. \qquad\; n=0,1,2,\cdots
\end{split}
\end{equation}
\noindent
As it can be seen, the fit for $\mathcal{G}=50$ approaches the 
result presented in Eq.~\eqref{MassLiterature}, while the 
fit for $\mathcal{G}=1$ is completely different. 

Although it is not considered the backreaction of the dilaton on 
the metric, the condensate is shown to be important for 
the dynamics of hadron formation in the dual field
theory. 
\noindent

It is worth  mentioning that the spectrum tends to a continuum as the parameter 
$\mathcal{G}$ goes to zero. In such a limit,
it is recovered the problem of a massive scalar field in an AdS background with a constant dilaton.

In the sequence of this work we explore the
finite-tem\-per\-a\-ture effects on the mass spectrum due to the 
presence of the gluon condensate (parameter  $\mathcal{G}$).

\section{Melting of scalar mesons}
\label{Sec:ScalarMesonsFiniteT}

\subsection{The finite-temperature holographic QCD model}

To take into account the finite-temperature effects on the 
dual field
theory, we need to consider a black hole on the gravity side.
The standard static black-hole solutions include a horizon
function, $f(z)$, in a metric of the form \eqref{ThermalAdSMetric}. 
The background geometry we consider here is described by
\begin{equation}\label{BHMetric}
\begin{split}
 ds^2=& g_{mn}dx^{m}dx^{n}\\
 =& e^{2A(z)}\left(-f(z)dt^2+\frac{1}{f(z)}dz^2+
 \delta_{ij}dx^{i}dx^{j}\right),
\end{split}
\end{equation}
where $f(z)=1-z^4/z_{h}^4$, $\delta_{ij}$ is the Kronecker delta 
and the Latin indices run over the spatial transverse coordinates 
($i=1,2,3$). The domain of the holographic coordinate is 
$0\leq z\leq z_h$, where
$z_h$ indicates the position of the black-hole event horizon. The 
Hawking temperature of the black hole is given by
\begin{equation}\label{BHTemperature}
T=\Bigg{|}\frac{f'(z_h)}{4\pi}\Bigg{|}.
\end{equation}
According to the AdS/CFT dictionary, $T$ is also the
temperature of the dual thermal field theory.

It is possible to observe the parameter dependence 
of the confinement/deconfinement temperature (or critical temperature) 
for gluons by analyzing the free energy of the 
dual thermal field theory. What is more, 
the free energy is obtained 
by calculating the Euclidean on-shell action of 
the background fields. In the present case, the 
gravitational action for both cases, 
the thermal AdS spacetime \eqref{ThermalAdSMetric} and 
the black-hole spacetime \eqref{BHMetric}, 
are given by \cite{Herzog:2006ra}
\noindent
\begin{equation}\label{EqOSBackground}
\begin{split}
\mathcal{S}^{\scriptscriptstyle{\text{Thermal}}}
=-\frac{1}{2\kappa^2}\int\,d^5x\,\sqrt{\tilde g}\,e^{-\Phi}
\left(\widetilde{R}+\frac{12}{\ell^2}\right),\\
\mathcal{S}^{\scriptscriptstyle{\text{BH}}}
=-\frac{1}{2\kappa^2}\int\,d^5x\,\sqrt{g}\,e^{-\Phi}
\left(R+\frac{12}{\ell^2}\right),
\end{split}
\end{equation}
\noindent
where $\kappa$ is the gravitational coupling, and
$\tilde g_{mn}$ and $g_{mn}$ are the corresponding metrics.  
To get the Euclidean on-shell 
action, which is related to the 
free energy of the dual thermal field theory, we first
obtain the Einstein equations (in Ricci form) 
from \eqref{EqOSBackground},
\noindent
\begin{equation}
\widetilde R=-20/\ell^2, \qquad R=-20/\ell^2.
\end{equation}
\noindent
Then, substituting these results into 
Eq.~\eqref{EqOSBackground} we get (for the thermal phase)
\noindent
\begin{equation}\label{EqOSThermal}
\mathcal{S}^{\scriptscriptstyle{\text{Thermal}}}_{\text{on-shell}}
=\frac{4\,V_3}{\kappa^2\,\ell^2}\int_{0}^{\tilde \beta} dt
\int_{z_0}^{\infty}dz\,\sqrt{\tilde g}\,e^{-\Phi},
\end{equation}
\noindent
where $V_3$ is the three-dimensional transverse volume, 
$\tilde\beta$ corresponds to the period of the Euclidean time 
$t\sim t+i\tilde\beta$, 
and $z_0$ is the UV cutoff which lies close to the boundary. 
Moreover, doing the same procedure we 
obtain the on-shell action of the black-hole 
phase, 
\noindent
\begin{equation}\label{EqOSBH}
\mathcal{S}^{\scriptscriptstyle{\text{BH}}}_{\text{on-shell}}
=\frac{4\,V_3}{\kappa^2\,\ell^2}\int_{0}^{\beta} dt
\int_{z_0}^{z_h}dz\,\sqrt{g}\,e^{-\Phi},
\end{equation}
\noindent
where the Euclidean time belongs to the
interval $0\leq t\leq\beta$, and $z_0$ is the 
UV cutoff located close to the boundary. 
Furthermore, the temperature $T=1/\beta$ is related to 
the event-horizon coordinate $z_h$ through 
$\beta=\pi z_h$. 

\begin{figure}[!ht]
\centering
\includegraphics[width=8.2cm]{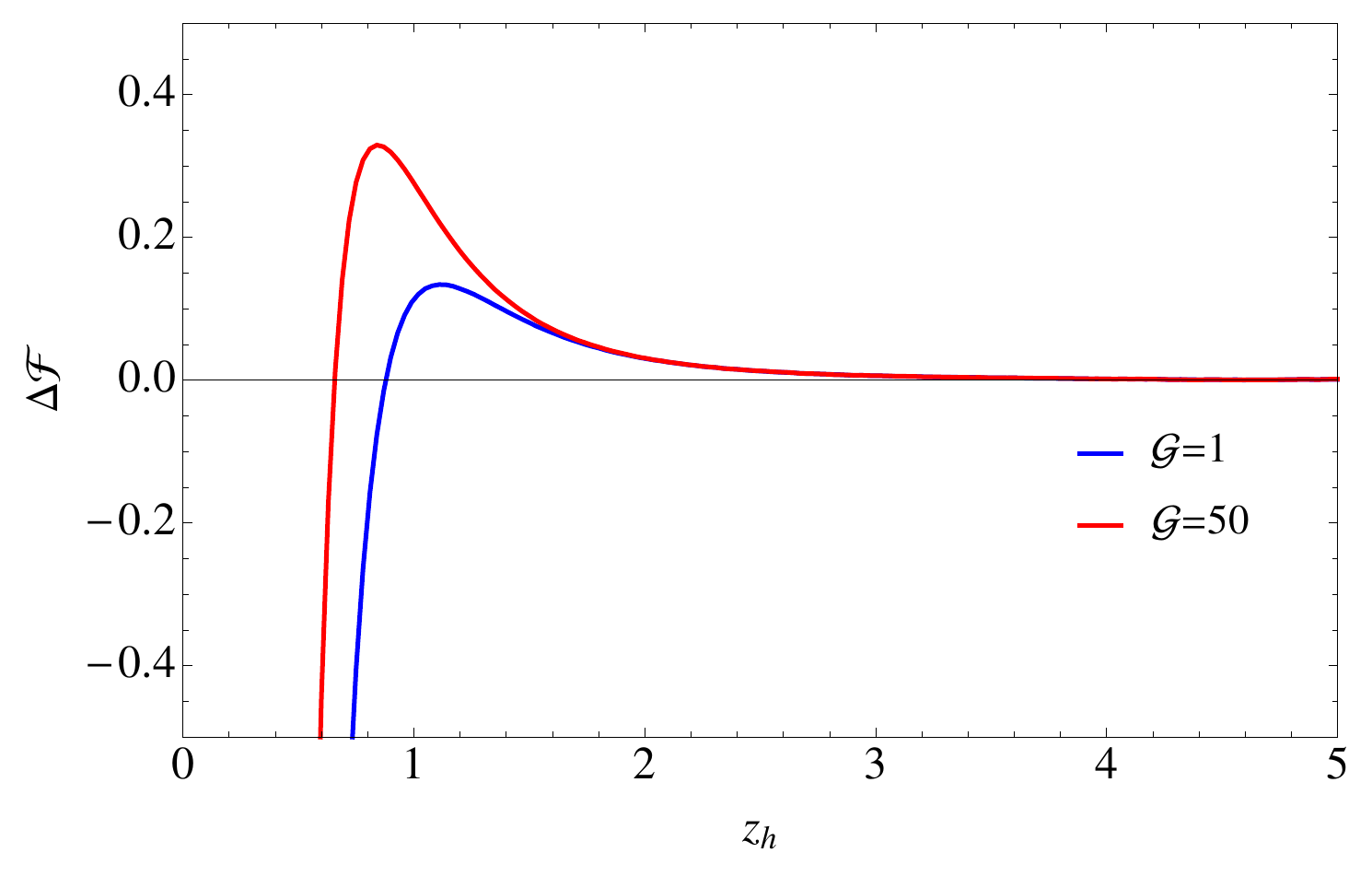} 
\caption{
The numerical 
results for $\Delta\mathcal{F}$ as a function of 
$z_h$ for $\mathcal{G}=1$ (blue line) and 
$\mathcal{G}=50$ (red line). We have set 
$C=1$ and $\kappa^2\,\ell^2=1$.}
\label{FigEQCD}
\end{figure}

The relation between the free 
energy $F$ and the Euclidean on-shell action 
is $\beta{F}=S_{\text{on-shell}}$. However, in both cases
the on-shell action blows up at the UV cutoff \cite{Herzog:2006ra}.
To avoid these divergences we are going
to use a prescription for which matters the difference of the
black hole and thermal phases.
Thus, the variation of the free energy density $\mathcal{F}=F/V_3$ is given by 
\noindent
\begin{equation}\label{EqEQCD}
\beta\Delta \mathcal{F}
=\lim_{z_0\to 0}
\frac{1}{V_3}\left(
\mathcal{S}^{\scriptscriptstyle{\text{BH}}}_{\text{on-shell}}
-
\mathcal{S}^{\scriptscriptstyle{\text{Thermal}}}_{\text{on-shell}}
\right).
\end{equation}
\noindent
Differently from the original soft-wall model \cite{Karch:2006pv}, here 
it is not possible to get an analytical expression for 
$\Delta\mathcal{F}$, so that
Eq.~\eqref{EqEQCD} must be solved 
numerically. To guarantee the
periodicity in time at the UV cutoff, we set 
$\tilde\beta=\beta\,\sqrt{f(z_0)}$ \cite{Herzog:2006ra}. In 
Fig.~\ref{FigEQCD} we display the numerical results 
obtained for 
$\mathcal{G}=1$ (blue line) 
and $\mathcal{G}=50$ (red line). 
The results in this figure show the 
parameter dependence of the difference in the free-energy density.
Furthermore, the confinement/deconfinement 
transition temperature is defined as the temperature 
for which the free-energy difference is zero. 
This means that we must solve the equation 
$\Delta\mathcal{F}(z_{h_{c}})=0$, where 
the solution $z_{h_{c}}$ is related to the 
critical temperature as $z_{h_{c}}=1/\pi T_c$. 
Moreover, from Fig.~\ref{FigEQCD} we observe that 
$z_{h_{c}}(\mathcal{G}=50)<z_{h_{c}}(\mathcal{G}=1)$, 
which means that 
$T_{c}(\mathcal{G}=50)>T_{c}(\mathcal{G}=1)$ and, 
consequently, 
the critical temperature is a function 
of the parameter $\mathcal{G}$. The inverse relation, i.e.,
the temperature dependence of the gluon condensate $\mathcal{G}(T)$, 
was already reported in the literature; 
for a discussion in QCD see, for instance, Ref.~\cite{Boyd:1996ex} 
and in holography Ref.~\cite{Andreev:2007zv}.

To complement this section we analyze the phase transition 
from the point of view of the degrees of freedom of 
the systems. A good thermodynamic variable to characterize 
this issue is the entropy. From the holographic dictionary 
we may determine the entropy of the dual field theory 
by calculating the entropy of the gravitational background 
with the use of the Bekenstein-Hawking formula, 
i.e., $S=\mathcal{A}/(4G_5)$, where $\mathcal{A}$ is the 
event horizon area.
As discussed in Ref.~\cite{BallonBayona:2007vp}, the number of
degrees of freedom of the deconfined phase is 
proportional to $N^2$ ($N$ is the number of colours), 
which means that $S\sim N^2$, 
while in the confined phase it is proportional to $N^0$, $S\sim N^0$. 
Therefore, the entropy function $S(N)$ has a phase 
transition because it is discontinuous at 
$T=T_c$. 

\subsection{Equations of motion for the scalar field}

The equation of motion for the scalar field $S(x,z)$ in the 
black-hole geometry is given by 
\noindent
\begin{equation}\label{ScalarFieldEqBH}
\begin{split}
& e^{-3A+\Phi}f\partial_z\left[e^{3A-\Phi}f\partial_z S(x,z)\right]
-\partial^{2}_{t} S(x,z)\\
&+f\left[\delta^{ij}\partial_{i}\partial_{j}
-m_X^2e^{2A}\right]S(x,z)=0.
\end{split}
\end{equation}
As in section \ref{Sec:ScalarMesons}, introducing the function $2B=3A-\Phi$ and
the tortoise coordinate $\partial_{r_{*}}=-f\,\partial_z$ in Eq.~\eqref{ScalarFieldEqBH},
and taking the Fourier transform
\begin{equation}\label{FourierTrans}
S(x,z)=\int \frac{d^4x}{(4\pi)^4}
e^{-i\omega t+i \mathbf k\cdot\mathbf{x}}S(k,x),
\end{equation}
we get
\begin{equation}\label{EqMotionFourierSpace}
e^{-2B}\partial_{r_{*}}\left[e^{2B}\partial_{r_{*}}S(k,z)\right]
+\left[\omega^2-f\left(q^2+m_X^2e^{2A}\right)\right]S(k,z)=0,
\end{equation}
where $q^2=\mathbf q \cdot\mathbf q=\mathbf k \cdot\mathbf k$ is the squared modulus of the 
spatial part of the four-momentum vector $k^{\mu}$. Finally, to get a Schr\"odinger-like equation,
we introduce the transformation $S=e^{-B}\psi$ and, after some simplifications, we find
\begin{equation}\label{SchroEqBH}
-\partial_{r_{*}}^2\psi(z)+V(z)\psi(z)=\omega^2\psi(z),
\end{equation}
where the potential is given by
\begin{equation}\label{TortPotential}
V(z)=\partial_{r_{*}}^2B+\left(\partial_{r_{*}}B\right)^2
+f(z)\left(q^2+m_X^2e^{2A}\right).
\end{equation}
\noindent
In the present case, it is helpful to write the explicit form of the tortoise 
coordinate in terms of the holographic coordinate $z$. Such a relation is given by
\begin{equation}\label{TortoiseCoord}
r_*=\frac{z_h}{2}\left[-\arctan\left(\frac{z}{z_h}\right)
+\frac{1}{2}\ln\left(\frac{z_h-z}{z_h+z}\right)\right].
\end{equation}
\noindent
This coordinate ranges from $r_{*}\to -\infty$  (for $z\to z_h$) to $r_{*}= 0$ (for $z= 0$).
An useful form for the effective potential as a function of $z$ is given by
\begin{equation}\label{EffPotentialT}
V(z)=f(z)\left[\partial_{z}\left(f(z)\partial_z B\right)
+f(z)\left(\partial_z B\right)^2
+q^2+m_X^2e^{2A}\right],
\end{equation}
from which one immediately sees that the potential vanishes at the event horizon.

\subsection{Asymptotic solutions for the scalar-meson field}
\label{SubSec:AsymptSols}

Here we study the asymptotic solutions of 
the differential equation~\eqref{SchroEqBH}. 
Since the effective potential \eqref{EffPotentialT} vanishes at the horizon, because
$f(z_h)=0$, the solutions for $\psi$ at $z=z_{h}$ are of the plane-wave form
\begin{equation}\label{AsymptSolHorizon1}
\psi \sim \mathfrak{C}\, e^{-i\omega r_*}
+\mathfrak{D}\, e^{+i\omega r_*},
\end{equation}
where the first term is a purely ingoing wave,
while the second term corresponds to a purely outgoing wave.
However, in the neighbourhood of the event horizon the
wave functions can also be expressed as power series in $(1-z/z_h)$, i.e., in series of the form
$\psi^{(\pm)}(z) = \sum_{n} a_n^{\pm}\left(1-{z}/{z_h}\right)^{n\pm i\omega z_h^2/4}$.
After substituting these series into the differential equation
\eqref{SchroEqBH}, we get, up to second order, 
\begin{equation}\label{AsymptSolHorizon2}
\begin{split}
\psi^{(\pm)}(z)=&\left(1-\frac{z}{z_h}\right)^{\pm i\omega z_h^2/4}
\bigg[1+a_1^{(\pm)}\left(1-\frac{z}{z_h}\right)\\
&+a_2^{(\pm)}\left(1-\frac{z}{z_h}\right)^2+\cdots\bigg],
\end{split}
\end{equation}
\noindent
where the ellipses represent higher-order contributions, and
the coefficients are given by
\noindent
\begin{equation*}
\begin{split}
a^{(\pm)}_1=&\pm \frac{i\,3z_h\,\omega}{8}+
\frac{\left(3\,C^2+C\left(6G+C\,q^2\right)z_h^2
+3\,G^2\,z_h^4\right)}{2\left(C+G\,z_h^2\right)^2
\left(2\pm i\,\omega\,z_h\right)}\\
&-\frac{\,G\left(4\,C+q^2\right)\left(2\,C+G\,z_h^2\right)z_h^4}
{8\left(C+G\,z_h^2\right)^2\left(2\pm i\,\omega\,z_h\right)},\\
\end{split}
\end{equation*}
\begin{equation*}
\begin{split}
a^{(\pm)}_2=&
\frac{1}{64\left(C+G\,z_h^2\right)^4\left(4\pm i\,\omega\,z_h\right)}
\bigg[
-G^4z_h^8\bigg(120+24q^2\,z_h^2\\
&\pm 68\,i\,\omega z_h
-17\omega^2z_h^2\bigg)-4\,C\,G^3\,z_h^6\bigg(120+24q^2\,z_h^2\\
&+56G\,z_h^4\pm 68\,i\,\omega\,z_h-17\omega^2\,z_h^2\bigg)
-c^4\bigg(120+24q^2\,z_h^2\\
&+960\,Gz_h^4-256G^2z_h^8
\pm 68\,i\,\omega z_h-17\omega^2 z_h^2\bigg)\\
&+4C^3Gz_h^2\bigg(-120-24q^2z_h^2-408Gz_h^4+64G^2z_h^8\\
&\mp 68\,i\,\omega z_h+17\omega^2 z_h^2\bigg)
+2C^2G^4 z_h^4\bigg(-360-72q^2 z_h^2\\
&-448G z_h^4
+32G^2 z_h^8\mp 204\,i\,\omega z_h+51\omega^2 z_h^2\bigg)\bigg]\\
& -\frac{a^{(\pm)}_1}
{16\left(C+Gz_h^2\right)^2\left(4\pm i\,\omega z_h\right)}
\bigg[G^2 z_h^4\bigg(84+4q^2 z_h^2\\
&\pm 30\,i\,\omega z_h
-3\omega^2 z_h^2\bigg)\\
& +2C\,G z_h^2\left(84+4q^2 z_h^2+8G z_h^4\pm 30\,i\,\omega z_h
-3\omega^2 z_h^2\right) \\
&+C^2\left(84+4q^2 z_h^2+32G z_h^4\pm 30\,i\,\omega z_h
-3\omega^2 z_h^2\right)\bigg].
\end{split}
\end{equation*}
\noindent
It is worth mentioning that the leading terms in 
Eqs.~\eqref{AsymptSolHorizon1} and \eqref{AsymptSolHorizon2} are  identical.
This can be shown by substituting the tortoise coordinate as a function of 
$z$ into Eq. \eqref{AsymptSolHorizon1} and expanding the resulting function 
in power series of $(1-z/z_h)$. We prefer to work with 
Eq.~\eqref{AsymptSolHorizon2} instead of Eq.~\eqref{AsymptSolHorizon1} in the 
forthcoming sections.

Now we turn attention to the solutions near the boundary $z=0$. 
The power series expansion of the two independent solutions may be written as
\begin{align}
\psi^{\scriptscriptstyle{(1)}}(z)=&z^{3/2}\left[1
+b_2\,z^2
+b_4\,z^4
+b_6\,z^6+\cdots\right],
\label{NormalSol}
\\
\psi^{\scriptscriptstyle{(2)}}(z)=&z^{-1/2}\left[1
+c_2\,z^2
+c_4\,z^4
+c_6\,z^6+\cdots\right]\nonumber\\
&+2d\psi^{\scriptscriptstyle{(1)}}\ln{\left(\frac{z}{z_h}\right)},
\label{NonNormalSol}
\end{align}
\noindent
where the ellipses stand for higher order terms, and the coefficients are
given by 
\begin{equation}\label{CoeffBoundarySol}
\begin{split}
b_2&=\frac{1}{8}\left(q^2-\omega^2\right),\qquad
b_4=\frac{3}{8z_h^4}
+\frac{1}{24z_h^2}b_2\left(q^2-\omega^2\right),
\\
b_6&=\frac{G^2}{8\,C}+\frac{11}{16z_h^6}b_2
+\frac{1}{48z_h^4}\left[-q^2+b_4\left(q^2-\omega^2\right)\right],
\\
d&=\frac{1}{4}\left(q^2-\omega^2\right),\qquad
c_2=\frac{3d}{2z_h^2}-\frac{1}{z_h^4\left(q^2-\omega^2\right)},
\\
c_6&=\frac{G^2}{4\,C}+\frac{3\,c_2}{8z_h^6}+\frac{17d}{48z_h^6}
-\frac{5d}{1152z_h^2}\left(q^2-\omega^2\right)^2
-\frac{q^2}{24z_h^4}.
\end{split}
\end{equation}
\noindent
Notice that $\psi^{(1)}$ is a normalizable wave function while $\psi^{(2)}$ is not,
since it diverges at the boundary.    

In order to calculate the spectral function associated to the scalar mesons 
(in Sec.~\ref{Sec:SpectralFunc}), we follow the same procedure applied in 
Refs.~\cite{Miranda:2009uw, Mamani:2013ssa}. First, we write the near-horizon 
ingoing and outgoing solutions, $\psi^{(-)}$ and $\psi^{(+)}$, as a 
linear combination of the wave functions 
close to the boundary,
\noindent
\begin{equation}\label{AsympHoriAsBoundSols}
\psi^{\scriptscriptstyle{(\pm)}}=
\mathfrak{A}^{\scriptscriptstyle{(\pm)}}
\psi^{\scriptscriptstyle{(2)}}+
\mathfrak{B}^{\scriptscriptstyle{(\pm)}}
\psi^{\scriptscriptstyle{(1)}},
\end{equation}
\noindent
where the coefficients are functions of the wave number and frequency. Second,
we may also write the near-boundary solutions
as a linear combination of the near-horizon solutions, 
\noindent
\begin{equation}\label{AsympBoundAsHoriSols}
\begin{split}
\psi^{\scriptscriptstyle{(1)}}
=&\mathfrak{C}^{\scriptscriptstyle{(1)}}
\psi^{\scriptscriptstyle{(-)}}
+\mathfrak{D}^{\scriptscriptstyle{(1)}}
\psi^{\scriptscriptstyle{(+)}}\,,\\
\psi^{\scriptscriptstyle{(2)}}
=&\mathfrak{C}^{\scriptscriptstyle{(2)}}
\psi^{\scriptscriptstyle{(-)}}
+\mathfrak{D}^{\scriptscriptstyle{(2)}}
\psi^{\scriptscriptstyle{(+)}}\,.
\end{split}
\end{equation}
\noindent
Doing so, we realized that there is a relation be\-tween 
Eqs. \eqref{AsympHoriAsBoundSols} and 
\eqref{AsympBoundAsHoriSols} which 
can be written in a matrix form (for details about this relation, see 
Refs.~\cite{Miranda:2009uw, Mamani:2013ssa}). To get the spectral function
associated to the mesons, we need to know some of the foregoing coefficients
($\mathfrak{A}^{\scriptscriptstyle{(-)}}$ and $\mathfrak{B}^{\scriptscriptstyle{(-)}}$,
say) and so, in the sequence of this work, we find numerically the values of these 
coefficients.

\subsection{Analysis of the finite-temperature effective potential}
\label{SubSec:EffectivePotT}

In this subsection we develop a careful analysis
of the effective potential \eqref{EffPotentialT}
and study how it varies with the parameters of the model.
First of all, let us write the potential explicitly as a 
function of the dilaton field $\Phi$, the holographic coordinate $z$
and the black-hole temperature $T$,
\begin{equation}\label{EffPotentialT2}
\begin{split}
V=&\frac{3}{4z^2}+\left[\frac{3}{2z}
+\frac{\Phi'(z)}{4}\right]\Phi'(z)-\frac{\Phi''(z)}{2}+q^2+\\
&\bigg[\frac32-z^2 q^2-\left(1+\frac{z}{2}\Phi'(z)\right)z\Phi'(z)
+z^2\Phi''(z)\bigg]z^2\pi^4 T^4\\
&-\frac{1}{2}\bigg[\frac{9}{2}+\left(1
-\frac{z}{2}\Phi'(z)\right)z\Phi'(z)
+z^2\Phi''(z)\bigg]z^6\pi^8 T^8.
\end{split}
\end{equation}
\noindent

Now it is helpful to consider some particular cases, beginning by
the zero-temperature potential. From Eq.~\eqref{EffPotentialT2},
it is clear that we recover the potential \eqref{EffPotential}
with the additional term $q^2$ for $T=0$,
\noindent
\begin{equation}
V(z){\Big |}_{T=0}=\frac{3}{4z^2}+\left[\frac{3}{2z}
+\frac{\Phi'(z)}{4}\right]\Phi'(z)-\frac{\Phi''(z)}{2}+q^2.
\end{equation}
\noindent
Thus, by choosing the dilaton as being a quadratic function
of the holographic coordinate $z$, we recover previous results
from the literature
(see, e.g, Refs. \cite{Colangelo:2008us, Bartz:2016ufc}).

Now let us turn off the gluon condensate by taking $G=0$.
In doing so, the dilaton field \eqref{DilaEq} reduces
to a constant, i.e., $\Phi(z)=\phi_0$, and the
potential~\eqref{EffPotentialT2} becomes
\noindent
\begin{equation}
V(z){\Big |}_{G=0}=\frac{3}{4z^2}+q^2
+\left(\frac{3}{2z^2}-q^2\right)z^4\pi^4 T^4
-\frac{9}{4}z^6\pi^8 T^8.
\end{equation}
\noindent
This is the same potential as that of a pure AdS 
black hole metric
with a massive scalar field.

\begin{figure*}[!ht]
\centering
\includegraphics[width=7.3cm]{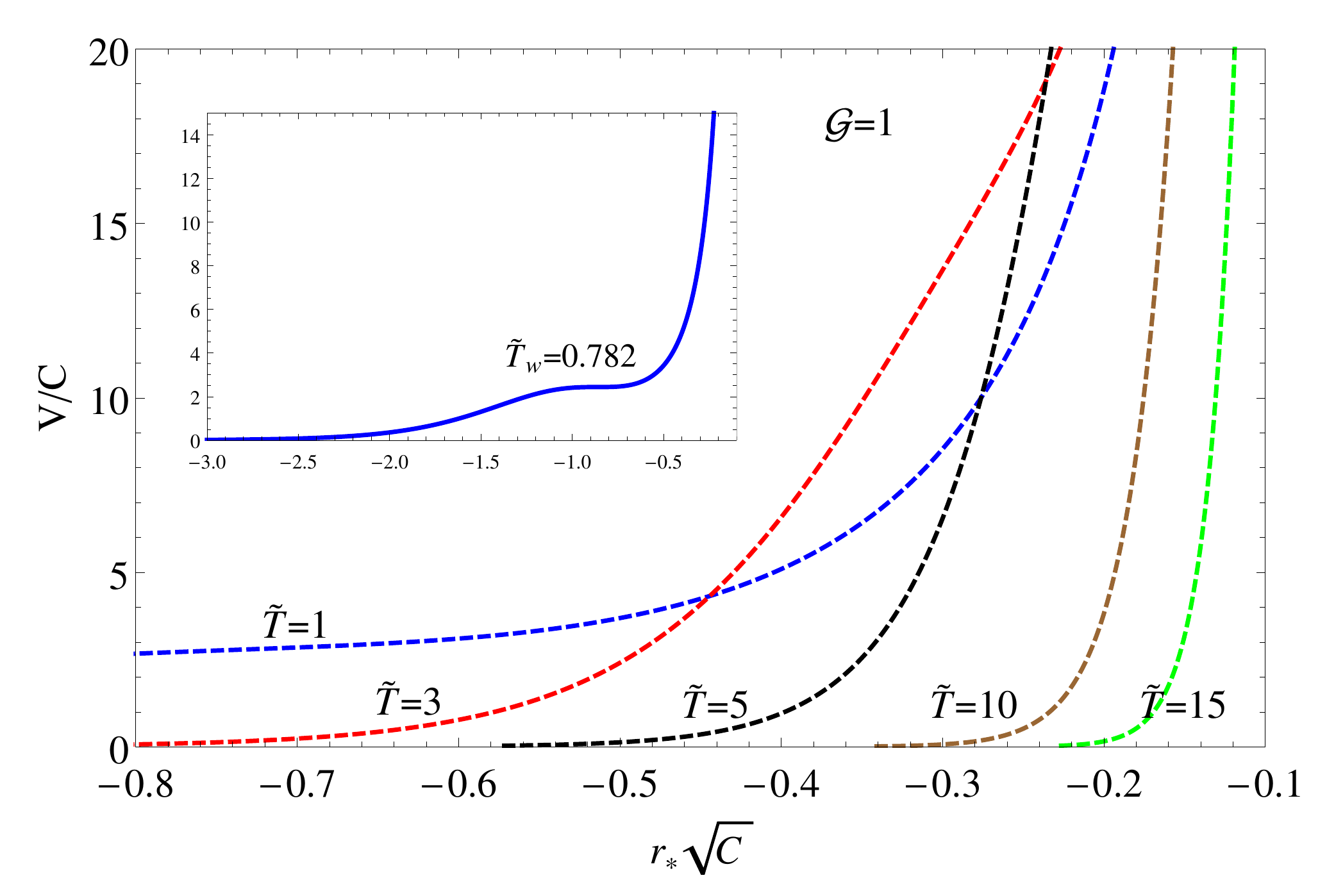}
\hspace{4em}
\includegraphics[width=7.3cm]{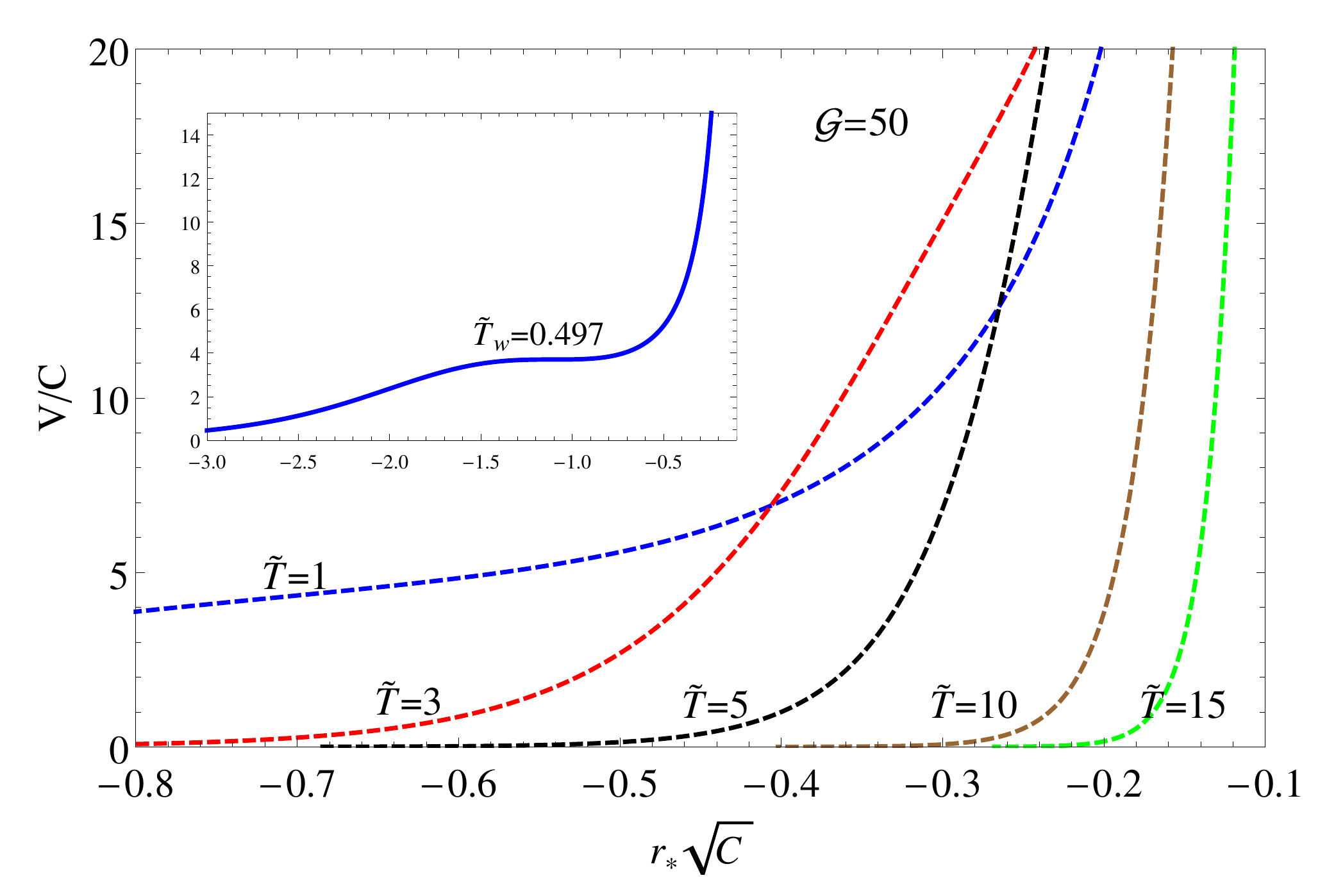}
\caption{Effective potential with zero wavenumber for $\mathcal{G}=1$ (left panel)
and $\mathcal{G}=50$ (right panel) in the regime of intermediate
and high temperatures. We also show the potential for $\widetilde{T}=\widetilde{T}_w$ in the inset.
}\label{PotentialHigh}
\end{figure*}

To implement the numerical analysis of the effective 
potential, it is convenient to normalize all parameters of the model in order to 
make them dimensionless. 
The temperature is normalized by the confinement energy scale as $\widetilde{T}=\pi T/\sqrt{C}$
\cite{Miranda:2009uw, Mamani:2013ssa}, and the wavenumber 
is normalized by the temperature,
$\bar{q}=q/\pi T$. We introduce again the dimensionless 
gluon condensate
$\mathcal{G}=G/C^2$.  The tortoise coordinate is normalized
as $\sqrt{C}\,r_{*}$ and the effective potential by $V/C$. 
For a better visualization of the graphs of the potential, we separate 
the analysis in two regimes:
one for low temperatures ($\widetilde{T}<1$) and another for intermediate 
and high temperatures ($\widetilde{T}\geq 1$). In 
Fig.~\ref{PotentialHigh} we show the results
for intermediate and high temperatures,  
and two values of the dimensionsless condensate, 
$\mathcal{G}=1$ (left panel) and $\mathcal{G}=50$ 
(right panel). These values were chosen arbitrarily and are the same
that we used to calculate the spectrum of the scalar mesons at zero temperature 
presented in Table \ref{Tab:01}.
 
By comparing both figures, we 
observe that the effect of the dimensionless
gluon condensate is more relevant for intermediate
temperatures; see, for example, the results for $\widetilde{T}=1$.
The results for high temperatures are less sensitive to 
this parameter, as can be seen by comparing the results for $\widetilde{T}=15$.

Another interesting result is the temperature for which the potential 
well starts to form, as is shown in the small box of Fig.~\ref{PotentialHigh}.
This temperature, that we call as $\widetilde{T}_w$, depends on
the value of $\mathcal{G}$, such that 
$\widetilde{T}_w=0.782$ for $\mathcal{G}=1$ is greater than
$\widetilde{T}_w=0.497$ for $\mathcal{G}=50$. As is shown 
in both figures, for $\widetilde{T}>\widetilde{T}_w$ the presence of bound states
is not expected due to the absence of a potential well.
The potential well starts to form for temperatures smaller than
$\widetilde{T}_{w}$. To finish this case,
it is noticed the temperature effect in deforming
the potential \eqref{EffPotential}, as can be easily
seen in the expression \eqref{EffPotentialT2}.
For high temperatures, the term $T^8$ becomes leading 
and its effect is visualized in Fig.~\ref{PotentialHigh}.

\begin{figure*}[!ht]
\centering
\includegraphics[width=7.3cm]{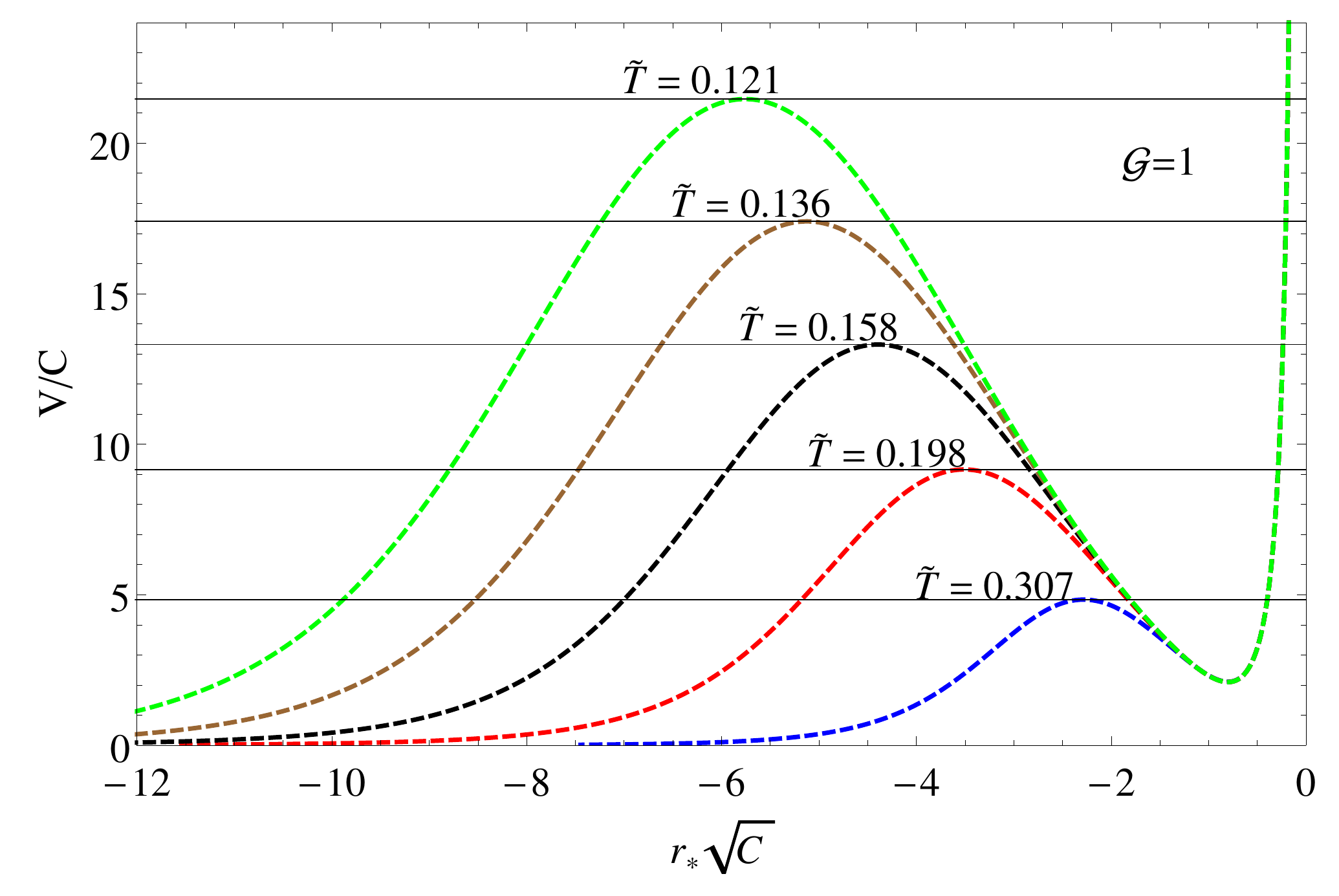}
\hspace{4em}
\includegraphics[width=7.3cm]{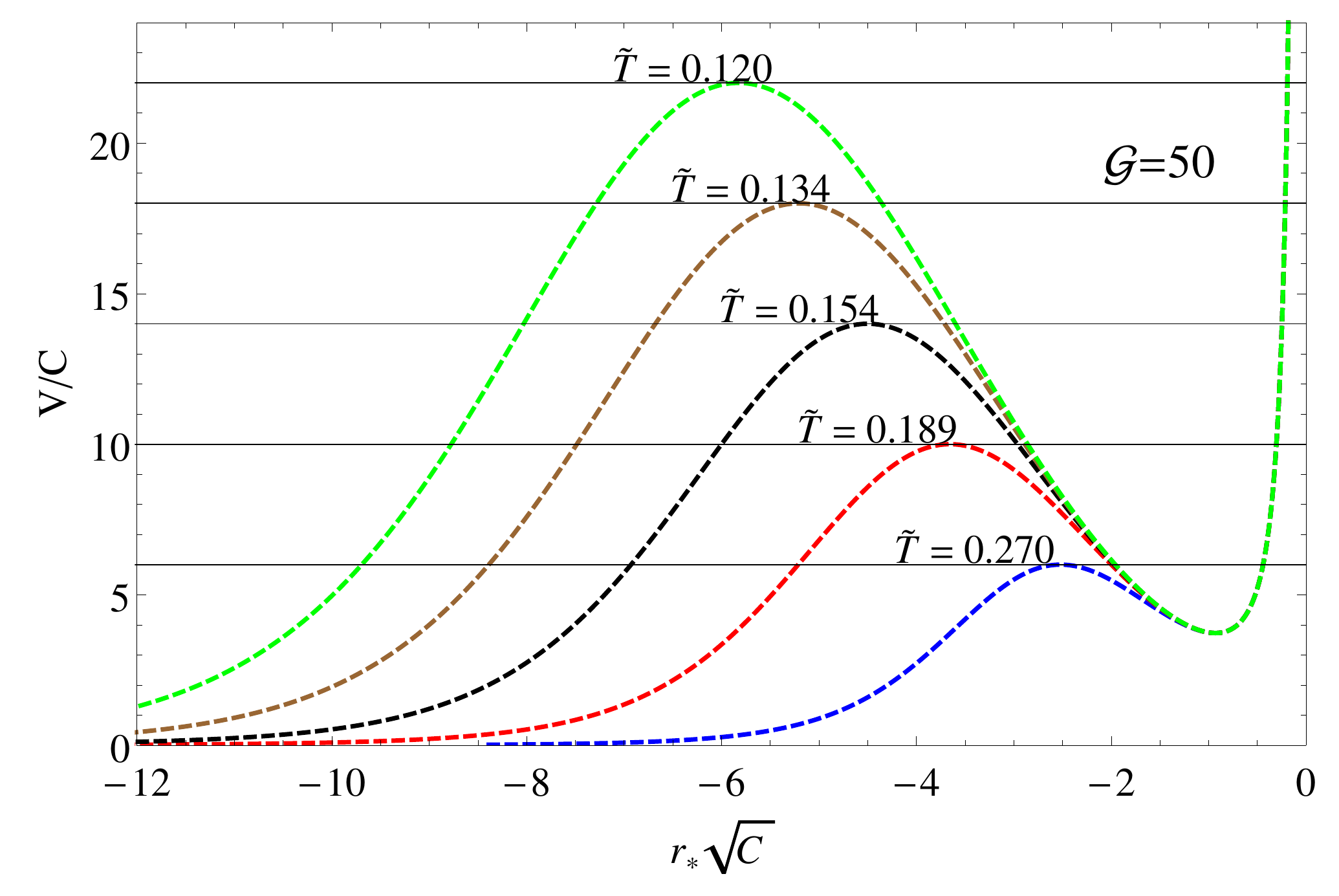}
\caption{Potential with zero wavenumber for
low temperatures for $\mathcal{G}=1$ (left panel)
and $\mathcal{G}=50$ (right panel). It is evident
how the potential is deformed with the temperature.
}\label{PotentialLow}
\end{figure*}

The curves for the potential at low tem\-per\-a\-tures are 
displayed in  
Fig. \ref{PotentialLow}. In this regime the potential 
presents a well and a barrier, with the height of the barrier depending on the
temperature.
Here we present the results for selected
values of temperature such that the maximum
value of the potential is equal to the mass
of the zero-temperature spectrum displayed in Table \ref{Tab:01}
and plotted as horizontal lines in this figure.
We did this analysis for $\mathcal{G}=1$ (left panel)
and $\mathcal{G}=50$ (right panel).
Our intention is always to compare with the results at zero 
temperature obtained in Sec.~\ref{Sec:ScalarMesons}.
From Fig.~\ref{PotentialLow} we see that the potential well
becomes deeper
with the decreasing of the temperature, and this
behaviour is the same
obtained for scalar glueballs and vector mesons in 
Refs.~\cite{Miranda:2009uw, Mamani:2013ssa}. For example, 
it is possible to have five bound states if the temperature
is smaller than $\widetilde{T}=0.121$ (on the left
panel of this figure).
But since the width of the potential is finite,
these bound states
have a finite lifetime, and the way they decay 
depends on the energy they have.
For a fixed value of temperature,
the higher exited states decay faster that the lower exited states.
A careful observation of Fig.~\ref{PotentialLow} shows how the 
potential \eqref{EffPotential} is deformed in the regime of
low temperatures. 

It is known from special relativity that $k_{\mu}k^{\mu}=-m^2_s$. In the case of a wavevector
with a vanishing spatial part, this result reduces to
$\omega^2=m^2_s$. This means that, at zero temperature, the frequency is
equal to the mass and the potential is shown in Fig.~\ref{Potential}.
However, in the finite-temperature case,
the Lorentz symmetry is broken and such a relation is no longer valid,
i.e., $k_{\mu}k^{\mu}\neq -m^2_s$.
As a consequence of it, the frequency acquires an imaginary
part that is related to the lifetime of the bound states.
This result is supported by the form of the potential in Fig.~\ref{PotentialLow}.

From these results is more evident the conclusion
that the dimensionless gluon condensate has a strong
influence on the low exited states and the higher exited 
states are less sensitive to this parameter as can be seen
by comparing the temperatures in both
panels in Fig.~\ref{PotentialLow}.

To finish the analysis of the effective potential we write
the potential close the boundary in its asymptotic form
\noindent
\begin{equation}
V=\frac{3}{4z^2}+\frac{6\,G^2}{C}z^4
+\cdots\left(\frac{3}{2z^2}+8Gz^2+\cdots\right)z^4\pi^4 T^4+\cdots,
\end{equation}
\noindent
where the ellipses represent higher order contributions
in the temperature and holographic coordinate. Or as
a functions of the tortoise coordinate this result 
becomes
\noindent
\begin{equation}\label{PotentialTortoise}
\begin{split}
V=&\frac{3}{4r_*^2}+\frac{6\,G^2}{C}r_{*}^4+\cdots\\
&+\left(\frac{9}{5r_{*}^2}
-\frac{24\,G^2}{5C}r_{*}^4+\cdots\right)r_{*}^4\pi^4 T^4
+\cdots,
\end{split}
\end{equation}
\noindent
as before, the ellipses represent subleading contributions. 
In this result
is more evident the contribution of the temperature to deform
the potential. Differently from the zero temperature case,
if we set $\mathcal{G}=0$ we still have bound states at 
finite temperature, the last term in 
Eq.~\eqref{PotentialTortoise} guarantee 
the existence of a potential well.

\subsection{Retarded Green function}

To find the real-time response in the dual field theory, we need
the retarded Green function. The AdS/CFT dictionary allows us
to find the correlation functions in the boundary field theory in
its Euclidean version \cite{Gubser:1998bc, Witten:1998qj}. 
A prescription to find two-point correlation functions was proposed in 
Ref. \cite{Son:2002sd}, and such a prescription is 
equivalent to the on-shell action re-normalization strategy as
developed in \cite{Skenderis:2002wp} (see also the references therein). 
It is important to point out that the scalar field $S(x,z)$ satisfies specific boundary conditions:  
Dirichlet at the AdS boundary and incoming wave at the horizon. 
These two boundary conditions guarantee that the poles
of the retarded Green function are precisely the black-hole QNM spectrum.
Here we obtain the two-point function following Ref.~\cite{Son:2002sd}.

Let us start by writing the action \eqref{ScalarAction} in the form
\begin{equation}\label{BulkAction}
\begin{split}
\mathcal{S}=&\int dx^5\sqrt{-g}e^{-\Phi}
S(x,z)\bigg[\frac{e^{\Phi}}{\sqrt{-g}}
\partial_{m}\left(
e^{-\Phi}\sqrt{-g}g^{mn}\partial_{n}\right)\\
&-m_X^2\bigg]S(x,z)
-\int dx^4 e^{-\Phi}\sqrt{-g}
g^{zz}S(x,z)\partial_{z}S(x,z)\bigg{|}_{z_0}^{z_h},
\end{split}
\end{equation}
where $z_0$ is a point close to the boundary and $z_h$ is
the position of the event horizon.
Since the equation of motion is satisfied,
the first term in Eq.~\eqref{BulkAction} is zero.
Hence, the on-shell action reduces to the surface term 
\noindent
\begin{equation}\label{OSActionT}
\mathcal{S}_{\text{on-shell}}=-\int dx^4 e^{-\Phi}\sqrt{-g}\,
g^{zz}S(x,z)\partial_{z}S(x,z)\bigg{|}_{z_0}^{z_h}.
\end{equation}
\noindent
After introducing the Fourier transform \eqref{FourierTrans} and 
decomposing the field as $S(k,z)=S_0(k)S_{k}(z)$, where $k$
is the four-momentum, the on-shell action \eqref{OSActionT} can be written as
\noindent
\begin{equation}\label{OnShellAct}
\mathcal{S}_{\text{on-shell}}=\int \frac{dk^4}{(2\pi)^4} 
S_0(-k)\mathcal{F}(k,z)S_0(k)\bigg{|}_{z_0}^{z_h},
\end{equation}
\noindent
where 
\begin{equation}\label{FluxFunct}
\mathcal{F}(k,z)=-e^{-\Phi}\sqrt{-g}\,g^{zz}
S_{-k}(z)\partial_{z}S_{k}(z).
\end{equation}
\noindent
This is similar to the
result obtained in Ref.~\cite{Son:2002sd}, with the main differences
being the presence of the dilaton and the fact that \eqref{FluxFunct}
is related to a massive field in the bulk.

Now the asymptotic solutions for the field $S_k(z)$, which satisfies an equation of motion
in Fourier space identical to Eq.~\eqref{EqMotionFourierSpace}, need to be found.
Here we also write the generic asymptotic expansion of the solution
for a massive scalar bulk field close to the 
boundary \cite{Skenderis:2002wp},
\noindent
\begin{equation}\label{AsymptGenSol}
S_{k}(z)=z^{4-\Delta}\left(1+\cdots\right)
+z^4\left(1+\cdots\right),
\end{equation}
\noindent
where $\Delta$ is the conformal dimension of the operator dual
to the scalar field $S$, 
and the ellipses mean subleading terms. Then, it is possible to
write the field as 
$S_{k}(z)=z^{4-\Delta}f_{k}(z)$, where the 
function $f_{k}(z)$ satisfies the condition
\noindent
\begin{equation}\label{UnitaryCond}
\lim_{z_0\to 0}f_k(z_0)=1.
\end{equation}
\noindent
After replacing Eq.~\eqref{AsymptGenSol} into Eq.~\eqref{FluxFunct}
and using Eq.~\eqref{UnitaryCond} it follows
\noindent
\begin{equation}\label{FluxFunctMod}
\mathcal{F}(k,z)=-e^{-\Phi}\sqrt{-g}\,g^{zz}\,
z^{4-\Delta}
f_{-k}(z)\partial_{z}\left[z^{4-\Delta}f_{k}(z)\right].
\end{equation}
\noindent
We use this result to obtain the retarded Green function following the
prescription of Ref.~\cite{Son:2002sd},
\noindent
\begin{equation}\label{eq:inutil2}
G^{R}(k)=-2\mathcal{F}(k,z).
\end{equation}
\noindent

In order to get the asymptotic solution for the bulk scalar field
$S(k,z)$, we use the results of the asymptotic solutions
close to the boundary, Eqs. \eqref{NormalSol} and \eqref{NonNormalSol},
together with the relations
\noindent
\begin{equation}\label{eq:inutil}
\begin{split}
S(k,z)&=e^{-B}\psi(k,z)=\\
&e^{-(3A+\Phi)/2}
\left[\mathfrak{A}^{\scriptscriptstyle{(-)}}
(\omega,\mathbf q)\psi^{(2)}(z)
+\mathfrak{B}^{\scriptscriptstyle{(-)}}
(\omega,\mathbf q)\psi^{(1)}(z)\right].
\end{split}
\end{equation}
Now it is easy to get the explicit expression for $f_k(z)$ from the relation
$S(k,z)=S_{0}(z)\,z^{4-\Delta}\,f_{k}(z)$,
\begin{equation}
f_k(z)=z^{1/2}e^{-\Phi/2} \left(\psi^{(2)}(z)
+\frac{\mathfrak{B}^{\scriptscriptstyle{(-)}}
(\omega,\mathbf q)} {\mathfrak{A}^{\scriptscriptstyle{(-)}}
(\omega,\mathbf q)}\psi^{(1)}(z)\right),
\end{equation}
where we have used $\Delta=3$ and 
$S_0(k)=\mathfrak{A}^{\scriptscriptstyle{(-)}}
(\omega,\mathbf q)$ to guarantee
the condition \eqref{UnitaryCond}. 
Finally, by using Eqs.~\eqref{FluxFunctMod}, \eqref{eq:inutil2} and \eqref{eq:inutil},
it is obtained the Green function
\noindent
\begin{equation}\label{RetardGreenFunct}
\begin{split}
G^R(\omega,\mathbf q)=&2(3c_2+2d)\\
&+12\,d\log\left[\frac{z_0}{z_h}\right]
+6\frac{\mathfrak{B}^{\scriptscriptstyle{(-)}}
(\omega,\mathbf q)}
{\mathfrak{A}^{\scriptscriptstyle{(-)}}
(\omega,\mathbf q)}
+\cdots,
\end{split}
\end{equation}
\noindent
where the ellipses denote power corrections in $z_0$. After a
renormalization process \cite{Skenderis:2002wp} we take the limit $z_0\to 0$ 
to extract the finite part of \eqref{RetardGreenFunct}.
The imaginary part of such a result is related 
to the spectral function (SPF), which is given by
\noindent
\begin{equation}\label{SpectFunct}
\mathcal{R}(\omega,\mathbf q)=-2\,\text{Im}G^R(\omega,\mathbf q)
=-12\,\text{Im} \frac{\mathfrak{B}^{\scriptscriptstyle{(-)}}
(\omega,\mathbf q)}
{\mathfrak{A}^{\scriptscriptstyle{(-)}}
(\omega,\mathbf q)}.
\end{equation}
\noindent

It is worth mentioning that the retarded Green function
could be obtained using the prescription presented in Ref. \cite{Iqbal:2008by}.
The authors used the canonical momentum associated
to the massive scalar field to get this quantity, and 
showed that their result is consistent with the 
prescription presented in Ref.~\cite{Son:2002sd}, at least
for a massless scalar field. Here we also used
the prescription of the canonical momentum
for a massive scalar field and obtain
the same result as Eq.~\eqref{RetardGreenFunct}.
In the next subsection we present the numerical results for
the spectral function \eqref{SpectFunct} and an analysis of
its dependence with the temperature and dimensionless gluon condensate.

\subsection{Spectral function}
\label{Sec:SpectralFunc}

\subsubsection{General procedure}

Here we make a brief summary of the general procedure to obtain the SPFs in holographic QCD.
In the present case, the spectral function is given by Eq.~\eqref{SpectFunct}. The next step is 
to express the coefficients 
$\mathfrak{B}^{\scriptscriptstyle{(\pm)}}(\omega,\mathbf q)$ and
$\mathfrak{A}^{\scriptscriptstyle{(\pm)}}
(\omega,\mathbf q)$ in terms of the 
asymptotic solutions of $\psi$ close to the boundary, as 
obtained in Sec.~\ref{SubSec:AsymptSols}. The idea is to write the solutions 
and their derivatives as
\begin{equation}\label{eq:inutil3}
\begin{split}
\psi^{\scriptscriptstyle{(a)}}(z)
&=\psi^{\scriptscriptstyle{(-)}}(z)
\mathfrak{C}^{\scriptscriptstyle{(a)}}+
\psi^{\scriptscriptstyle{(+)}}(z)
\mathfrak{D}^{\scriptscriptstyle{(a)}},\quad\\
\partial_z\psi^{\scriptscriptstyle{(a)}}(z)
&=\partial_z\psi^{\scriptscriptstyle{(-)}}(z)
\mathfrak{C}^{\scriptscriptstyle{(a)}}+
\partial_z\psi^{\scriptscriptstyle{(+)}}(z)
\mathfrak{D}^{\scriptscriptstyle{(a)}}, 
\qquad (a=1,2),
\end{split}
\end{equation}
where $a=1$ ($a=2$) stands for the normalizable (non-nor\-mal\-izable)
solutions.  
In matrix form, Eq.~\eqref{eq:inutil3} reads

\begin{equation}\label{Matrix1}
\left(\begin{array}{cc}
\psi^{\scriptscriptstyle{(a)}}(z)\\
\partial_{z}\psi^{\scriptscriptstyle{(a)}}(z)
\end{array}\right)
=
\left(\begin{array}{cc}
\psi^{\scriptscriptstyle{(-)}}(z) \quad & \quad
\psi^{\scriptscriptstyle{(+)}}(z)\\
\partial_{z}\psi^{\scriptscriptstyle{(-)}}(z)\quad &
\quad \partial_{z }\psi^{\scriptscriptstyle{(+)}}(z)
\end{array}\right)
\left(
\begin{array}{cc}
\mathfrak{C}^{\scriptscriptstyle{(a)}}\\
\mathfrak{D}^{\scriptscriptstyle{(a)}}
\end{array}\right).
\end{equation}

The aim of this procedure is to get expressions for the 
coefficients $\mathfrak{D}^{\scriptscriptstyle{(a)}}$ and
$\mathfrak{C}^{\scriptscriptstyle{(a)}}$ as functions
of the asymptotic solutions. Hence, inverting the 
matrix product
\eqref{Matrix1} we get the desired result. To be more 
specific, we are looking for the ratio 
 \begin{equation}
\frac{\mathfrak{D}^{\scriptscriptstyle{(2)}}}
{\mathfrak{D}^{\scriptscriptstyle{(1)}}}
=\frac{\psi^{\scriptscriptstyle{(2)}}\partial_{z }\psi^{\scriptscriptstyle{(-)}}
-\psi^{\scriptscriptstyle{(-)}}\,
\partial_{z}\psi^{\scriptscriptstyle{(2)}}}
{\psi^{\scriptscriptstyle{(1)}}\partial_{z}\psi^{\scriptscriptstyle{(-)}}-
\psi^{\scriptscriptstyle{(-)}}\,
\partial_{z}\psi^{\scriptscriptstyle{(1)}}},
\end{equation}
since there is a connection between the coefficients 
$\mathfrak{D}^{\scriptscriptstyle{(1)}}$, 
$\mathfrak{D}^{\scriptscriptstyle{(2)}}$ and
$\mathfrak{B}^{\scriptscriptstyle{(-)}}$,
 $\mathfrak{A}^{\scriptscriptstyle{(-)}}$, given
by
\noindent
\begin{equation}
\frac{\mathfrak{B}^{\scriptscriptstyle{(-)}}}
{\mathfrak{A}^{\scriptscriptstyle{(-)}}}=
-\frac{\mathfrak{D}^{\scriptscriptstyle{(2)}}}
{\mathfrak{D}^{\scriptscriptstyle{(1)}}}.
\end{equation}
\noindent 
With the analytic expressions in hand, we use the asymptotic solutions
\eqref{NormalSol} and \eqref{NonNormalSol} as ``initial'' conditions
and solve numerically the Schr\"odinger-like equation~\eqref{SchroEqBH} from a point
close to the boundary $z=z_{\epsilon}$ up to a point close to the horizon $z_{nh}=z_h-z_{\epsilon}$,
where $z_{\epsilon}$ is a sufficiently small positive number, e.g., $z_{\epsilon}=0.001$.

\subsubsection{Numerical results}

Here we present and discuss some numerical results 
obtained following the general procedure ex\-pla\-i\-ned previously. 
Firstly, we calculate the spectral functions by 
setting $\mathbf q=0$. 
This means that the spatial components of the momentum are neglected,
so that the four-momentum is given by $k^{\mu}=(\omega,\mathbf 0)$. 
Hence, the bound states in the field theory do not present spatial displacement.

\begin{figure*}[!ht]
\centering
\includegraphics[width=7.3cm]{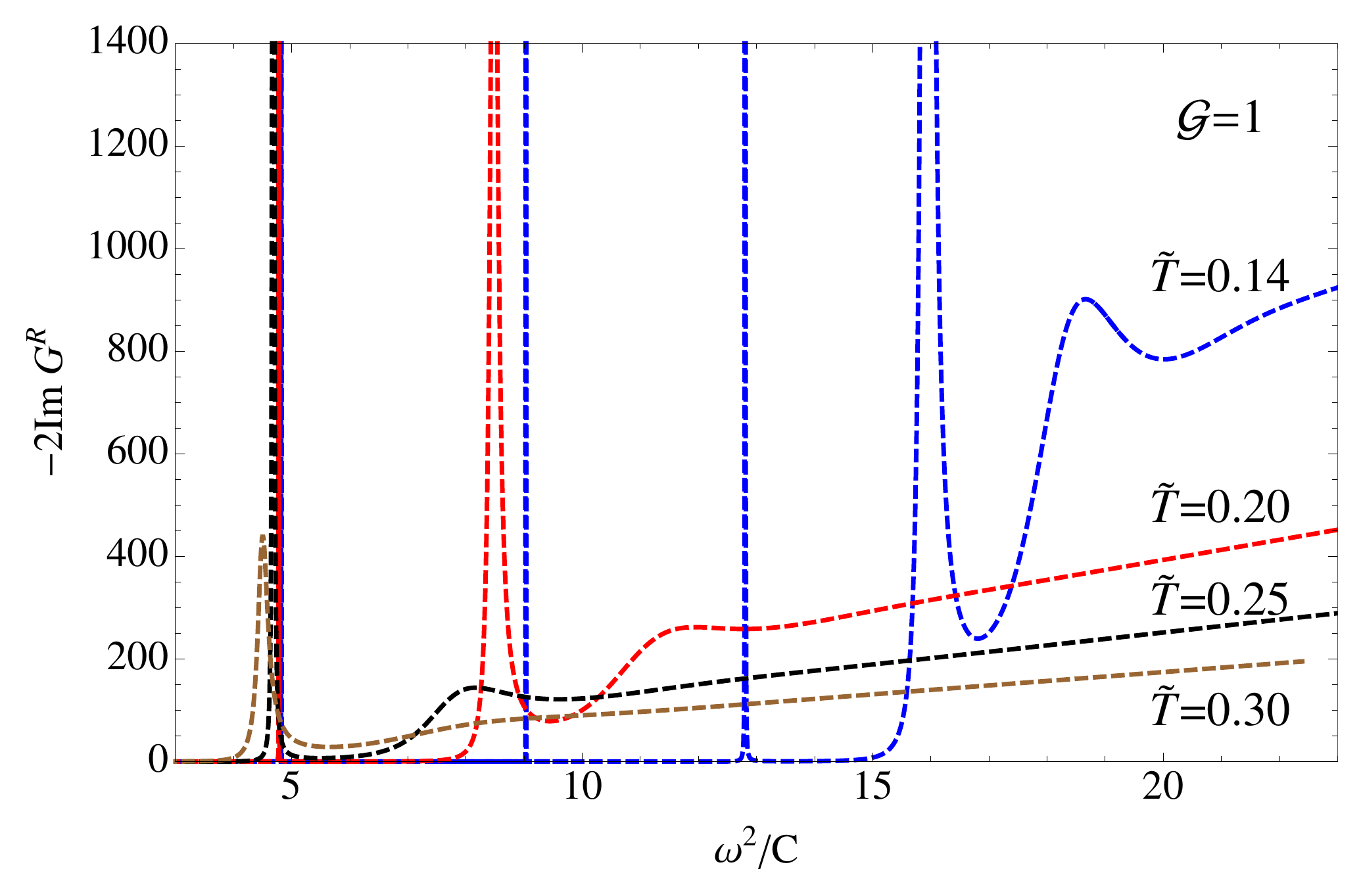}
\hspace{4em}
\includegraphics[width=7.3cm]{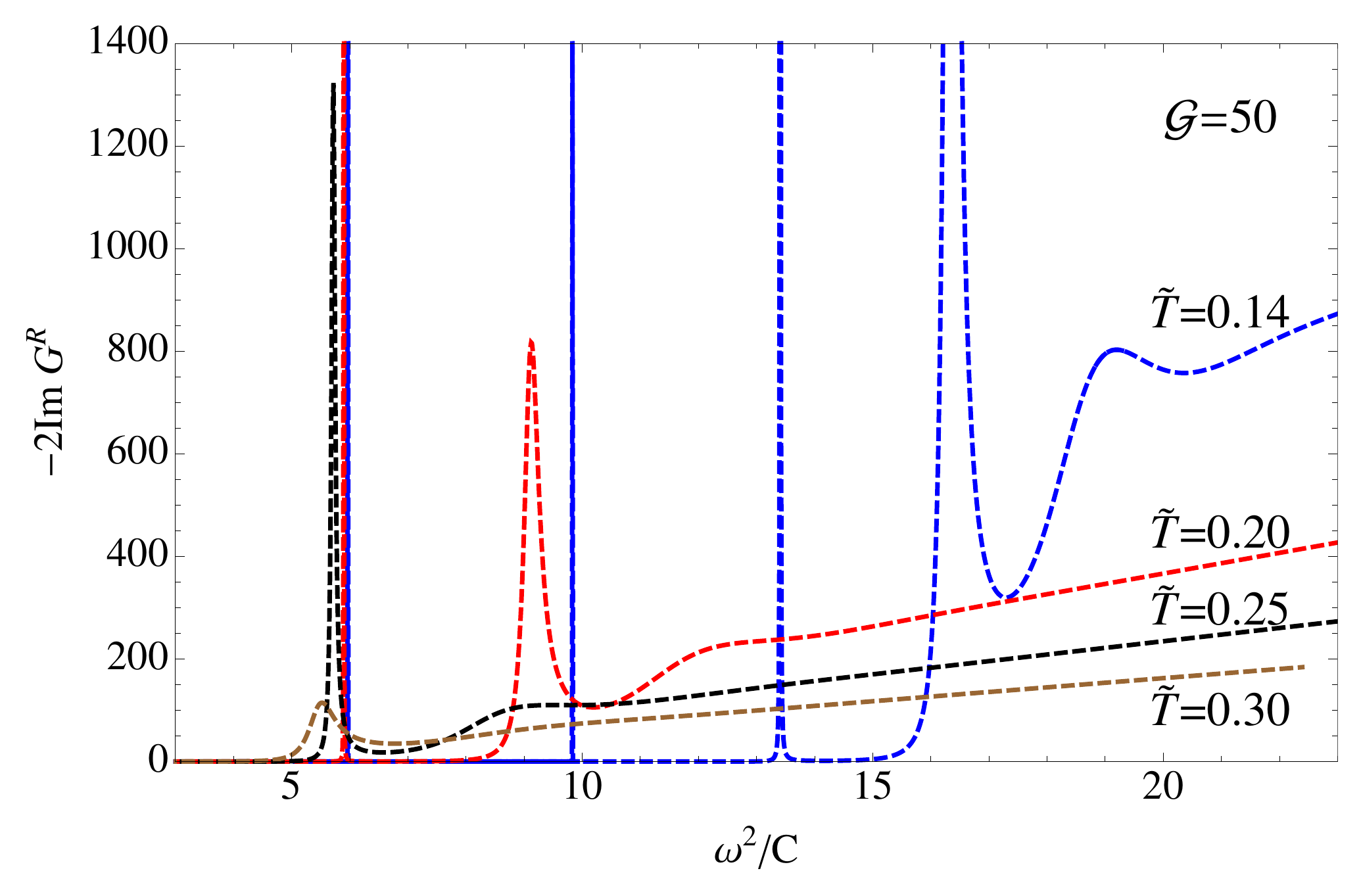}
\caption{Spectral functions with zero wavenumber in the
low-temperature regime for $\mathcal{G}=1$ (left panel)
and $\mathcal{G}=50$ (right panel). The number of peaks decreases with the
temperature in both cases.} 
\label{SpectralFuncLow}
\end{figure*}

Figure~\ref{SpectralFuncLow} shows the numerical results
for the spectral function in the low temperatures regime for 
$\mathcal{G}=1$ (left\- panel) and $\mathcal{G}=50$ (right panel).
Each curve in a given panel is drawn for a given temperature, as specified in the figure. 
The curves show several sharpen peaks, and the numerical data around each peak 
fit well a Breit-Wigner function \cite{Fujita:2009ca,Miranda:2009uw, Colangelo:2009ra,Mamani:2013ssa,Bartz:2016ufc}
\begin{equation}
\mathfrak{R}(\omega,\mathbf 0)=
\frac{\mathcal{A}\,\omega^{\,\mathfrak{b}}}
{\left(\omega-
{\omega}_{\scriptscriptstyle{0}}\right)^2
+\Gamma^{2}}\,,
\end{equation}
where $\omega_{\scriptscriptstyle{0}}$ is the position
of the peak, $\Gamma$ is the half-width of the curve around the peak, and
$\mathcal{A}$ and $\mathfrak {b}$ are constants that can 
be determined by fitting the numerical data in each case.
The peaks are interpreted as quasiparticle states.

The position of each curve peak in Fig.~\ref{SpectralFuncLow}
is related to the mass of the corresponding quasiparticle state, while
the half-width of 
the peak is related to the 
inverse of the quasiparticle lifetime. For instance,
in the case $\mathcal G=1$ the mass at zero temperature, 
cf. Table \ref{Tab:01}, for the fundamental state is
$m^2=4.84\,C$, and it is possible to get the position of the
first peak at $\omega^2_{\scriptscriptstyle{0}}=4.78\,C$ for 
the finite temperature $\widetilde{T}=0.2$.
Repeating the same comparison
in the case $\mathcal G=50$ we get $m^2=5.99\,C$ and  
$\omega^2_{\scriptscriptstyle{0}}=5.90\,C$ for 
$\widetilde{T}=0.2$. The difference between $m^2$ and 
$\omega_0^2$ depends 
on the temperature for fixed $\mathcal{G}$, i.e.,
$\Delta m^2(\widetilde{T}) =m^2-\omega^2_{\scriptscriptstyle{0}}$.
Note that this difference is positive for nonzero temperature.
However, at zero temperature the
spectral function reduces to delta functions with peaks
located at the values of the masses presented in Table
\ref{Tab:01}. In such a limit the difference is zero, i.e., 
$\Delta m^2(0)=0$. When the temperature is introduced the 
peaks acquire a width as it can be seen in Fig.\ref{SpectralFuncLow}. 
As the temperature increases the width of the peaks increases 
while the position of the peaks are shifted. We also see the 
number of peaks decreasing as the temperature increases, 
signalling the melting of the quasiparticle states.

In Ref.~\cite{Fujita:2009ca} the SPFs for scalar mesons were obtained 
by using the soft-wall model. The position of the 
first peak obtained for $\widetilde{T}=0.22$ is at 
$\omega^2_{\scriptscriptstyle{0}}=5.85\,C$. In order to 
compare to our results, we have calculated the SPFs 
for this temperature, but we do not show 
the corresponding
graphics here. For $\mathcal{G}=1$ the first peak is localized
at $\omega^2_{\scriptscriptstyle{0}}=4.75\,C$, while for
$\mathcal{G}=50$ it is localized at 
$\omega^2_{\scriptscriptstyle{0}}=5.85\,C$. From these 
results we observe, again, that the results of the 
soft-wall model are obtained when the value of the 
parameter $\mathcal{G}$ is large, just as we 
observed when the zero-temperature spectrum was calculated, 
cf. Sec.~\ref{SubSec:Spectrum}.

\begin{figure}[!ht]
\centering
\includegraphics[width=8.2cm]{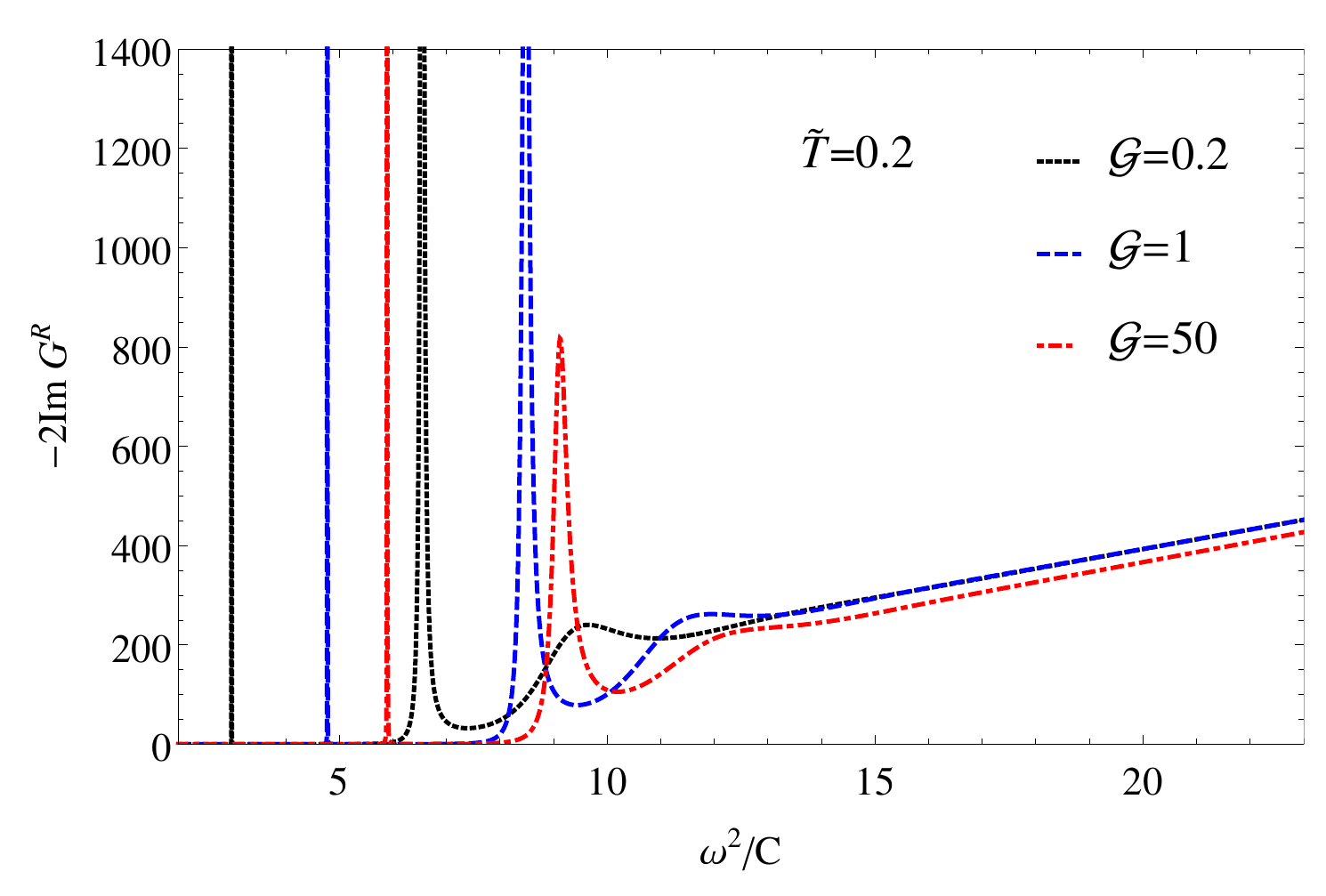}
\caption{
The spectral functions with zero wavenumber for 
$\mathcal{G}=0.2$, $\mathcal{G}=1$ and $\mathcal{G}=50$. 
These results were obtained for $\widetilde{T}=0.2$.
}\label{SpectralFuncCompar1}
\end{figure}

Additionally, 
the results shown in Fig.~\ref{SpectralFuncLow}
are in agreement with the results
obtained from the effective potential in 
Fig.~\ref{PotentialLow}. For instance, in the analysis
of the potential it was shown the possible 
existence of two quasiparticle states
for $\widetilde{T}=0.2$ in both cases,
$\mathcal{G}=1$ and $\mathcal{G}=50$. 
This is also supported by the SPFs 
where there are two peaks for the same temperature. 
The effect of the parameter $\mathcal{G}$ shows up in 
shifting the position, height, and width of the peaks 
(see Fig.~\ref{SpectralFuncCompar1}). In other words, the
position, height and width of the peaks are functions of 
$\widetilde{T}$ and $\mathcal{G}$ too, i.e.,
$\omega_{\scriptscriptstyle{0}}(\widetilde{T},\mathcal{G})$
and $\Gamma(\widetilde{T},\mathcal{G})$.

\begin{figure*}[!ht]
\centering
\includegraphics[width=7.3cm]{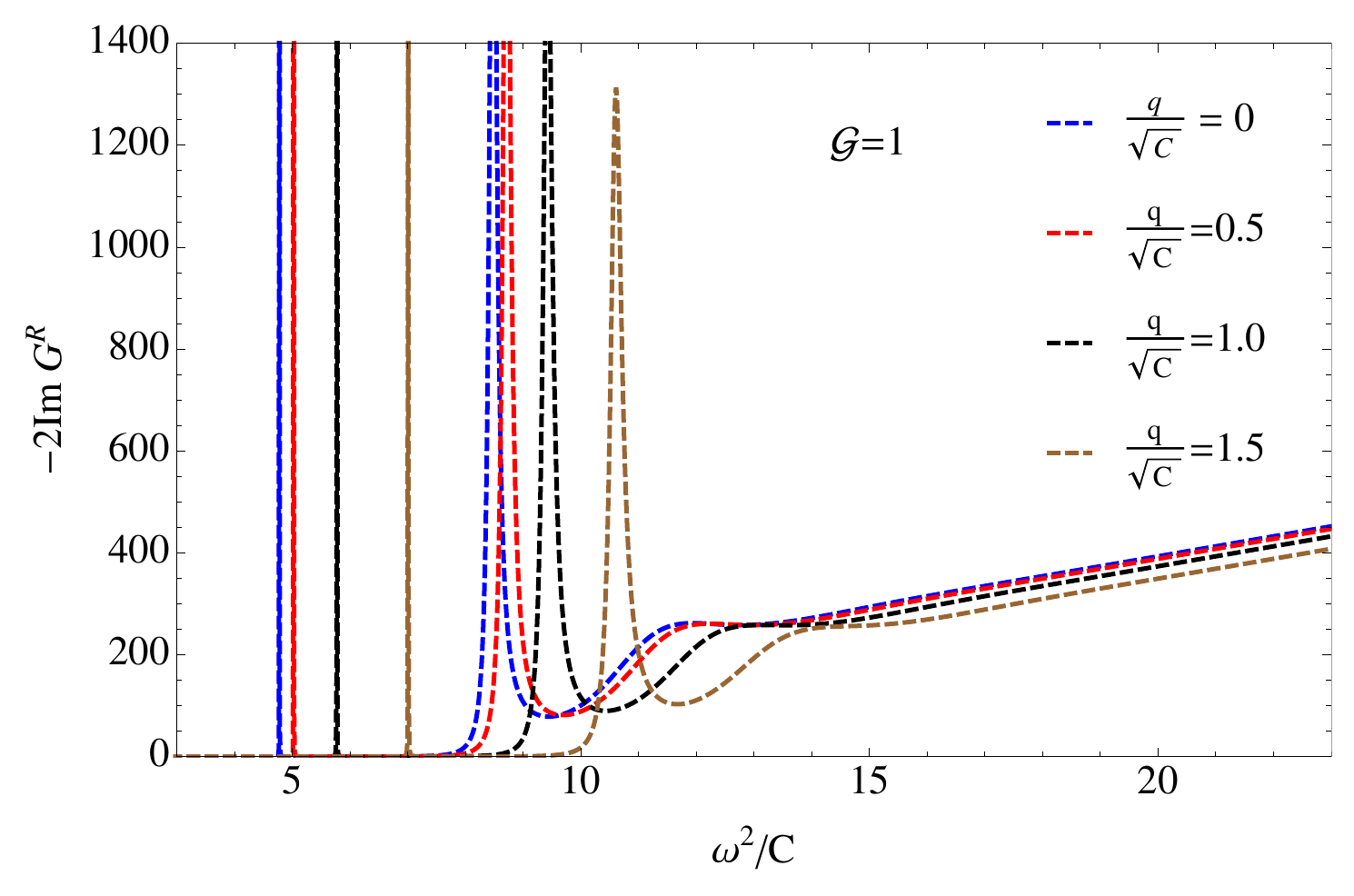}
\hspace{4em}
\includegraphics[width=7.3cm]{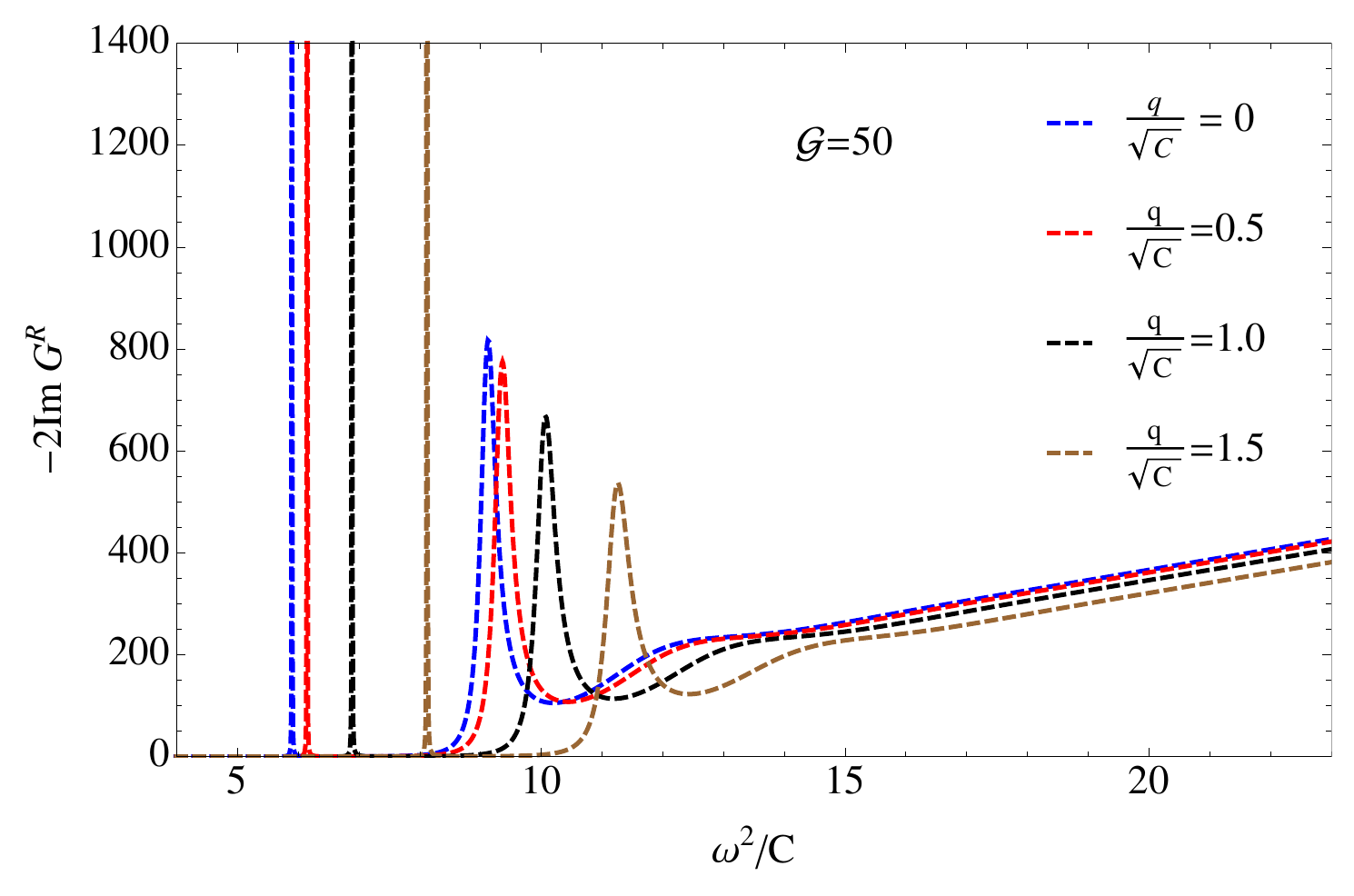}
\caption{Spectral functions 
with different wavenumber values for $\mathcal{G}=1$
(left panel) and $\mathcal{G}=50$  (right panel), in
both cases we set $\widetilde{T}=0.2$.
}\label{SpectralFuncMomentum}
\end{figure*}

It is noticed that the parameter $\mathcal{G}$ 
shifts $\omega_{\scriptscriptstyle{0}}$ to more energetic states, while the width of the peak increases.
To finish the analysis of the SPFs we consider a finite spatial momentum
in the above results, i.e., $\mathbf q\neq 0$. Hence, the quasiparticle states have spatial displacement and
therefore become more energetic. The results 
for this case are presented
in Fig.~\ref{SpectralFuncMomentum} for
$\mathcal{G}=1$ (left panel) and $\mathcal{G}=50$
(right panel). 
The addition of spatial momentum
shifts the position of the peaks to more energetic states, and
the half-widths also increase as the spatial momentum
increases. This means that more energetic states, i.e.,
quasiparticles with higher momenta,
have shorter lifetime and therefore they melt faster than
the low energy states.

\section{Quasinormal modes}
\label{Sec:QNMs}

In this section we present the results for the spectrum of black-hole 
quasinormal modes (QNMs) associated to the scalar field
$S(x,z)$. The field backreaction on the black-hole geometry is neglected, 
so that the scalar field is considered as a probe field.

\subsection{Power series method}

The standard procedure to use the 
power series method is to transform
the second-order eigenvalue problem,
characterized by a dependence in $\omega^2$,
into a differential equation which depends linearly in
$\omega$ \cite{Horowitz:1999jd}. To 
do that we introduce the
following transformation into the Schr\"odinger-like
equation~\eqref{SchroEqBH}
\noindent
\begin{equation}
\psi(z)=e^{-i\omega r_{*}}\varphi(z).
\end{equation}
\noindent
After replacing $dz/dr_{*}=-f(z)$ and simplifying, 
the following equation shows up
\noindent
\begin{equation}\label{FirstOrderProb}
f(z)\,\varphi''(z)
+\big[2i\omega+f^\prime(z)\big]\varphi^\prime(z)
-\frac{V(z)}{f(z)}\,\varphi(z)=0,
\end{equation}
\noindent
where $V(z)$ is given by Eq.~\eqref{EffPotentialT} and $\varphi(z)$ 
is a regular function in the interval $[0,z_h]$. 
As in previous sections, we
use the wave function behavior near the horizon to label the two independent
solutions of Eq.~\eqref{FirstOrderProb} by $\varphi^{(+)}$ (the outgoing wave)
and $\varphi^{(-)}$ (the ingoing wave).  
Now we write the 
solution for $\varphi^{(-)}$ as a power series
expansion around the horizon $z=z_h$,  
\noindent
\begin{equation}
\varphi^{\scriptscriptstyle{(-)}}(z)=
\sum_{n=0}^{\infty}\,a^{\scriptscriptstyle{(-)}}_{n}
\left(1-\frac{z}{z_h}\right)^{n},
\end{equation}
\noindent
where the coefficients $a_n$
are functions of the parameters of the model and, by convention, we set $a_0=1$. 
The spectrum of QN frequencies is 
obtained by imposing Dirichlet condition at the boundary, 
i.e., $\varphi^{\scriptscriptstyle{(-)}}(0)=0$, 
which is equivalent to solve the following 
recursive equation
\noindent
\begin{equation}
\sum_{n=0}^{\infty}\,a^{\scriptscriptstyle{(-)}}_{n}
(\omega,q,T,C,G)=0.
\end{equation}
\noindent
This method, first employed in Ref.~\cite{Horowitz:1999jd}, works
well for large AdS black holes, i.e., for high temperatures.
For low temperatures the numerical convergence gets poor, 
as it was previously noticed in holographic QCD models for 
scalar glueballs and vector mesons \cite{Miranda:2009uw,Mamani:2013ssa}.

\subsection{Breit-Wigner method}

As commented previously, it is known that the power series method is
not reliable to find QN frequencies in the 
regime of low temperatures. In such a regime, however, 
the form of the effective potential (cf. 
Sec.~\ref{SubSec:EffectivePotT}) allow us to 
apply a resonance method to obtain the black-hole QN frequencies. In this 
subsection we use 
the Breit-Wigner (resonance) method to calculate the 
frequencies associated to the scalar-field perturbations.
This method was applied for the very first time in Ref.~\cite{Berti:2009wx} 
to compute QNMs of small Schwarzschild-AdS black holes.
In the context of holographic QCD this 
numerical method was previously applied in Refs.~\cite{Miranda:2009uw,Mamani:2013ssa}.

Now we make a brief summary of the Breit$-$Wigner meth\-od; see
Refs.~\cite{Berti:2009wx,Miranda:2009uw,Mamani:2013ssa} 
and references therein for details. Firstly, we write the normalizable
solution $\psi^{\scriptscriptstyle{(1)}}$ of the Schr\"odinger-like equation
\eqref{SchroEqBH}, in the neighborhood of the horizon, as
\begin{equation}\label{equation223}
\begin{split}
\psi^{\scriptscriptstyle{(1)}}
&=\mathfrak{C}^{\scriptscriptstyle{(1)}}
e^{-i\omega r_*}+
\mathfrak{D}^{\scriptscriptstyle{(1)}} e^{+i\omega r_*}\\
&=
\alpha\,(\omega)\cos (\omega r_*)-
\beta\,(\omega) \sin (\omega r_*)\,.
\end{split}
\end{equation}
The ingoing-wave boundary condition at the horizon requires the vanishing of the coefficient  $\mathfrak{D}^{\scriptscriptstyle{(1)}}$. This means 
that, for $\omega$ close to the QN frequency, we might approximate it as 
$\mathfrak{D}^{\scriptscriptstyle{(1)}} \sim
(\omega-\omega_{\scriptscriptstyle{\text{QNM}}})$, where 
$\omega_{\scriptscriptstyle{\text{QNM}}}
=\omega_{\scriptscriptstyle{R}}-
i\,\omega_{\scriptscriptstyle{I}}$. Notice also that, from Eq.~\eqref{equation223},
$\mathfrak{C}^{\scriptscriptstyle{(1)}}
=\mathfrak{D}^{{\scriptscriptstyle{(1)}}\,*}$.
Taking this information into consideration we obtain
\begin{equation}\label{equation224}
\alpha^2+\beta^2=4\,\mathfrak{C}^{\scriptscriptstyle{(1)}}
\,\mathfrak{D}^{\scriptscriptstyle{(1)}}\approx
\text{const.}\times\left[(\omega-\omega_{\scriptscriptstyle{R}})^2+
\omega_{\scriptscriptstyle{I}}^2\right],
\end{equation}
where $\omega$ is a real parameter. Hence, the real par of $\omega$ can be obtained by minimizing 
Eq.~\eqref{equation224}. In turn, the 
imaginary part of the frequency
is obtained by fitting a parabola 
to Eq.~\eqref{equation224}. Alternatively it is 
possible to obtain the imaginary part of the frequency using the following
formulas \cite{Berti:2009wx,Miranda:2009uw,Mamani:2013ssa}
\begin{equation}\label{equation225}
\omega_{\scriptscriptstyle{I}}
=-\frac{\alpha}{\partial_{\omega}\beta}=
\frac{\beta}{\partial_{\omega}\alpha}.
\end{equation}
In the numerical analysis below, we take 
into consideration the value of the imaginary part obtained 
using these two procedures.

\subsection{Pseudo-spectral method}

The pseudo-spectral method \cite{boyd:2001} is used here to solve 
the linear eigenvalue problem \eqref{FirstOrderProb}. In this method,
the regular solution of Eq.~\eqref{FirstOrderProb} is expanded
in terms of
cardinal functions $C_j(z)$. To do that we first make
the substitution $\varphi(z)=(z/z_h)^{3/2}\, g(z)$, such that
$g(z)$ is a regular function in the interval $[0,z_h]$. The 
differential equation for $g(z)$, which follows from Eq.~\eqref{FirstOrderProb},
can be written in the form 
\noindent
\begin{equation}
\begin{split}
&\lambda_2(z,q,T,C,G)\,g''(z)+
\lambda_1(z,\omega,q,T,C,G)\,g'(z)\\
&+
\lambda_0(z,\omega,q,T,C,G)\,g(z)=0,
\end{split}
\end{equation}
\noindent
where $\lambda_{2}, \lambda_{1}$ and $\lambda_{0}$ are
linear polynomials in $\omega$. 
Now we expand the regular function 
$g(z)$ as
\noindent
\begin{equation}\label{PseudoEq}
g(z)=
\sum_{j=0}^{N}\,g(z_j)\,
C_j(2z/z_h-1),
\end{equation}
\noindent
where $C_j(2z/z_h-1)$ are the Chebyshev polynomials, and 
the collocation points are chosen to be the
Chebyshev-Gauss-Lobatto grid \cite{boyd:2001}: 
\noindent
\begin{equation}
z_j=\frac{z_h}{2}
\left[1-\cos\left(\frac{j\,\pi}{N}\right)\right].
\end{equation}
\noindent
By substituting \eqref{PseudoEq} into \eqref{FirstOrderProb} it gives
the matrix form of the eigenvalue problem, 
\noindent
\begin{equation}\label{EqEigenFreq}
(A+\omega\,B)\cdot\vec{g}=0,
\end{equation}
\noindent
where $A$ and $B$ are $(N+1)\times (N+1)$ matrices and $\vec{g}$ is a $(N+1)-$dimensional vector
with components $g_j=g(z_j)$, $j=0,\,1,...,N$. It 
is not difficult to implement numerically a code to solve 
Eq.~\eqref{EqEigenFreq}.

It is worth mentioning that the pseudo-spectral method inevitably leads to the 
emergence of spurious solutions that do not have any physical 
meaning. To eliminate the spurious solutions we use the 
fact that the relevant QN frequencies do not depend on the number of
Chebyshev polynomials being considered in Eq.~\eqref{PseudoEq}.

\subsection{Numerical results}

\begin{figure*}[!ht]
\centering
\includegraphics[width=7.3cm]{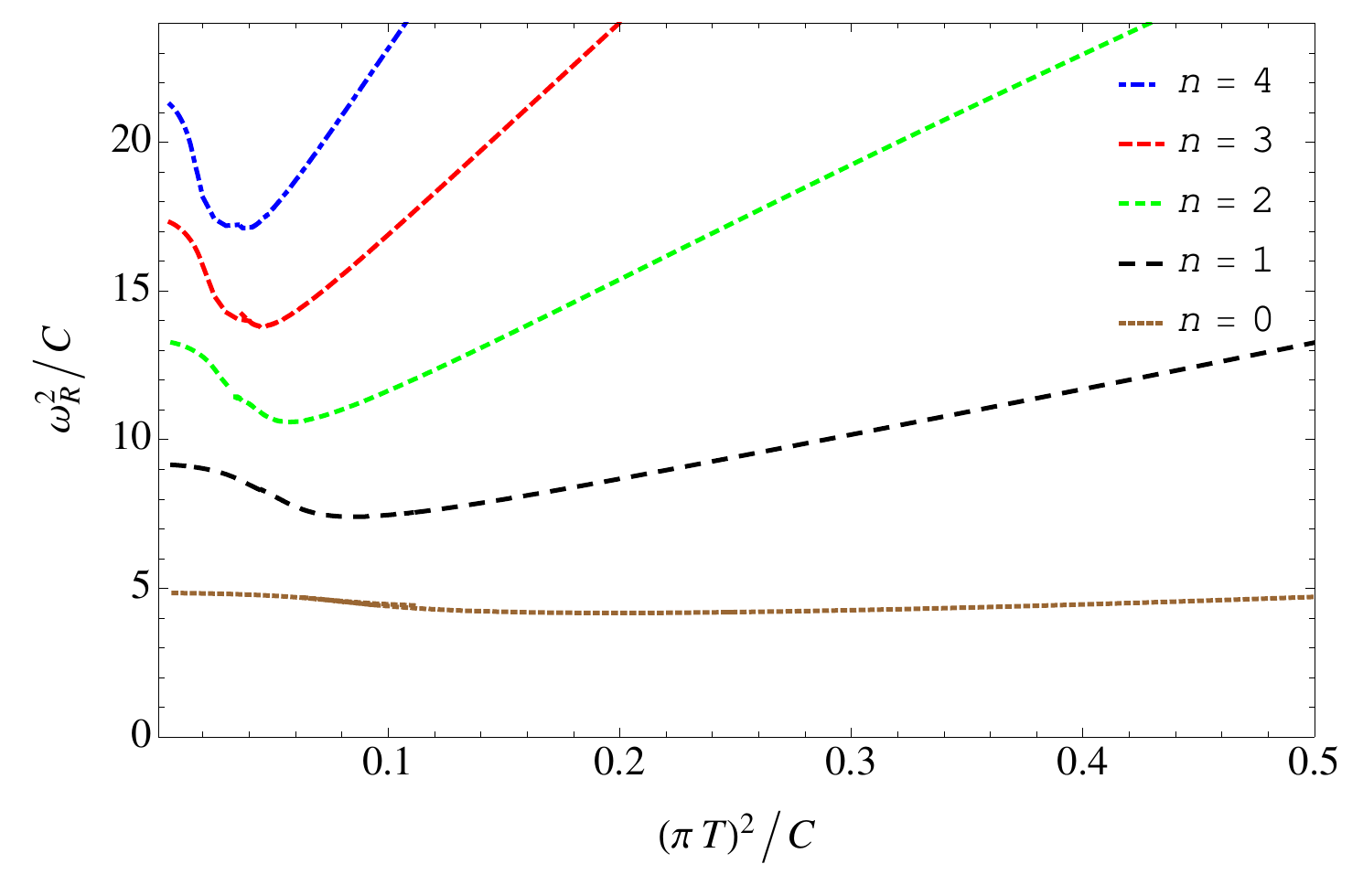}
\hspace{4em}
\includegraphics[width=7.3cm]{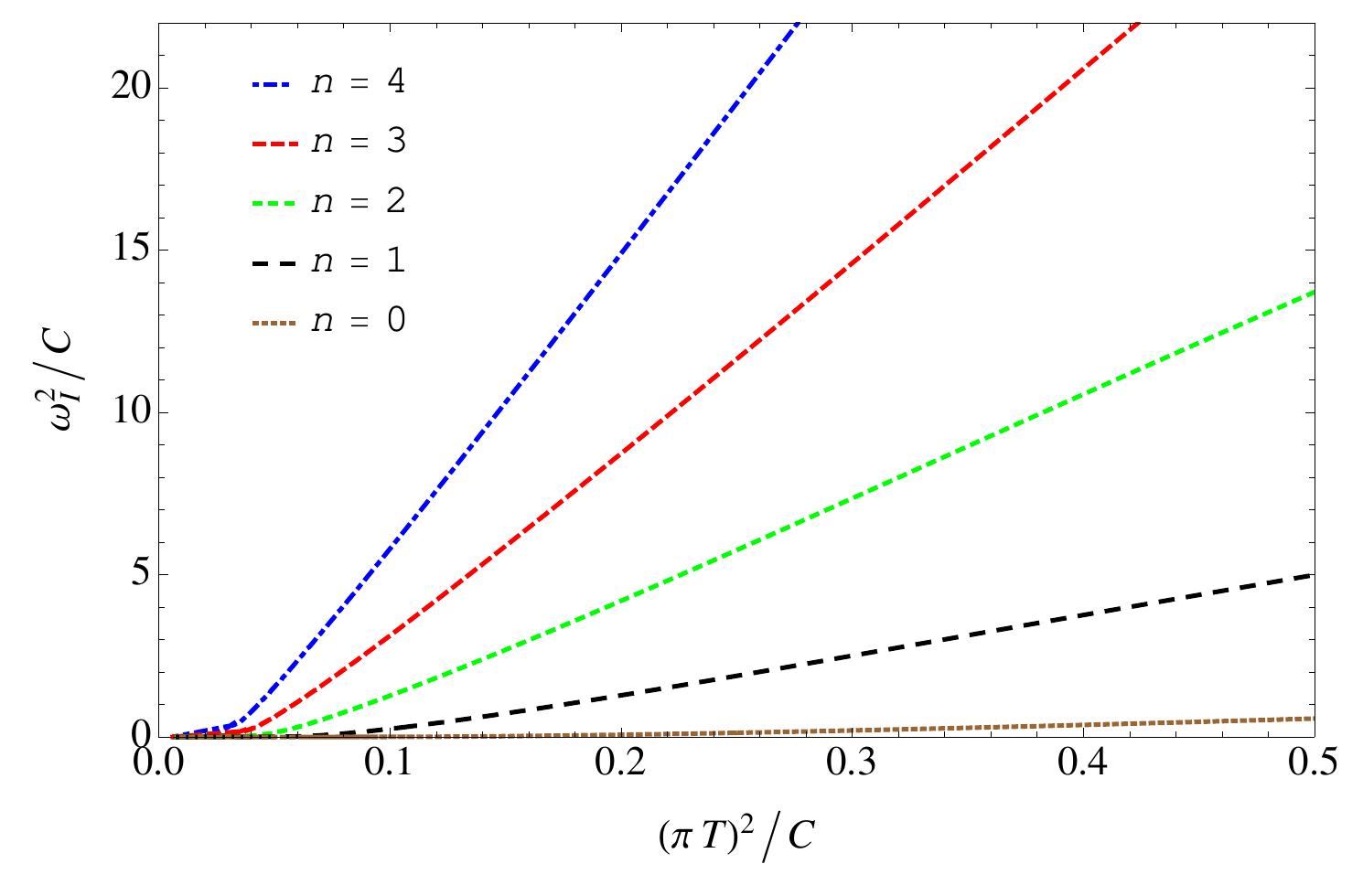}\\
\includegraphics[width=7.3cm]{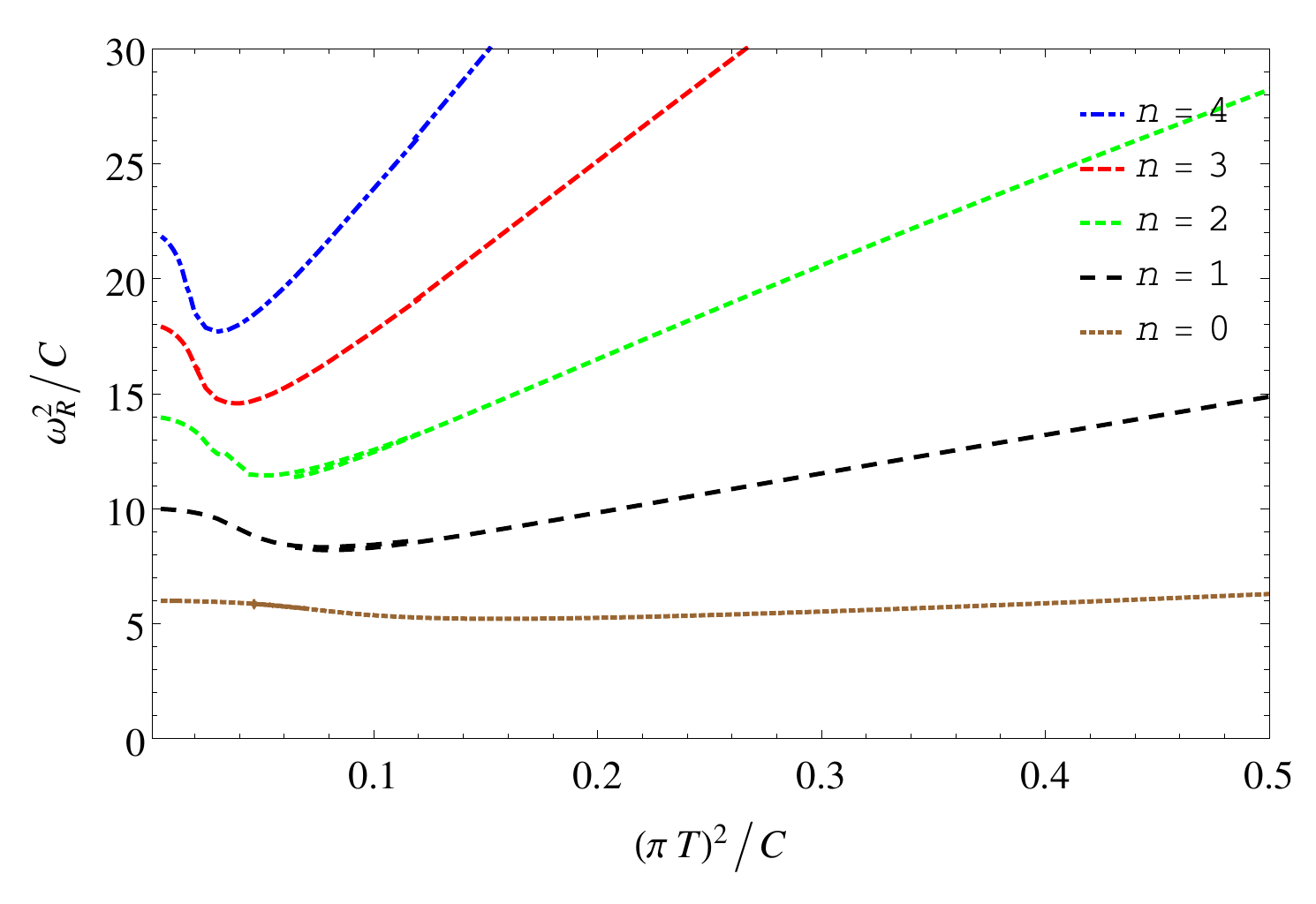}
\hspace{4em}
\includegraphics[width=7.3cm]{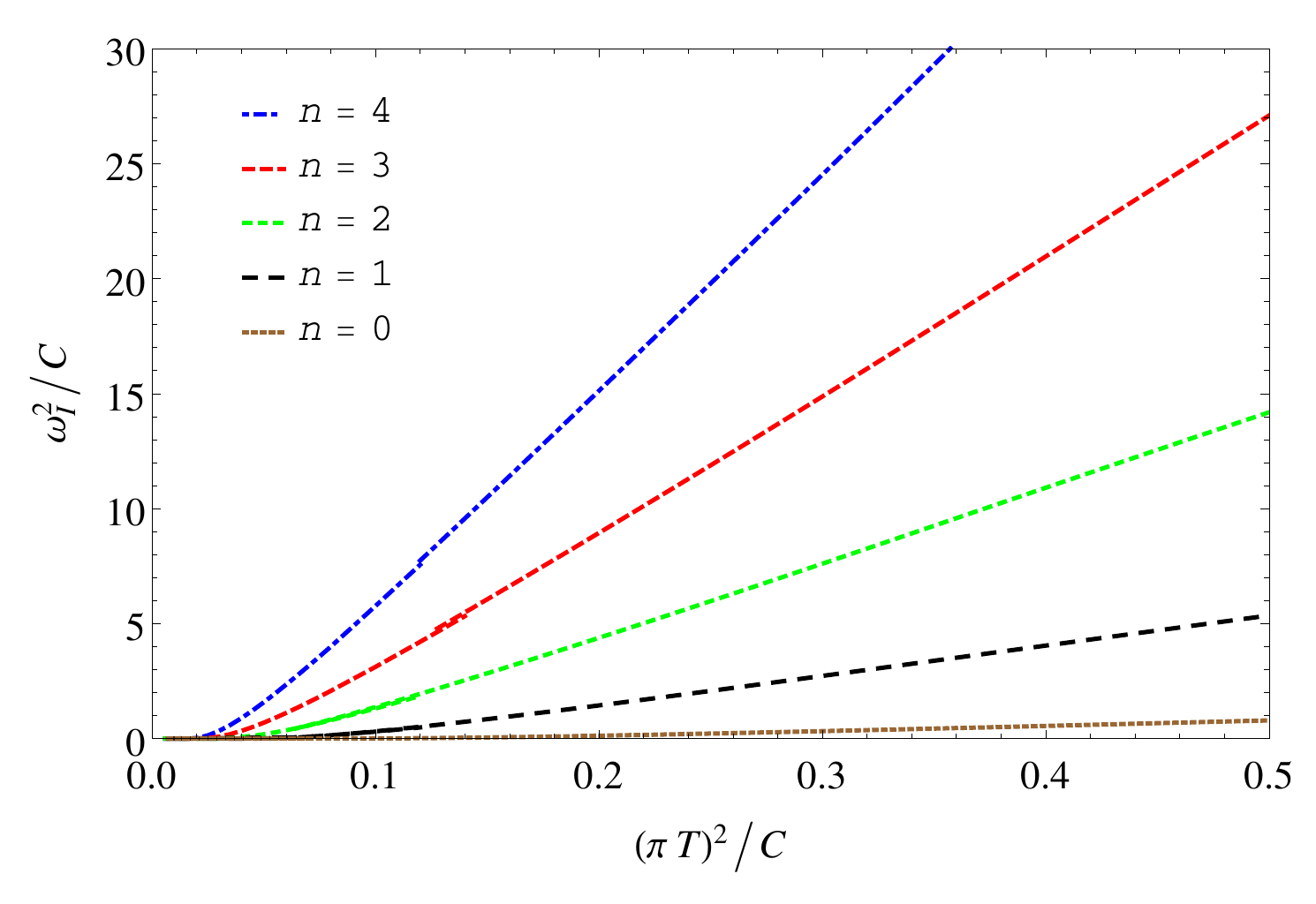}
\caption{The first five QN frequencies obtained 
using the Breit-Wigner, 
power series, and pseudo-spectral methods. The top panels 
show the results obtained for $\mathcal{G}=1$, 
while the bottom panels show the results for $\mathcal{G}=50$.
}\label{SpectrumG1G50}
\end{figure*}

Here we present and discuss the numerical results for the
QN frequencies obtained with the use of the foregoing methods.
Firstly, in Fig.~\ref{SpectrumG1G50} we display the overlap of 
the results obtained using the power series, pseudo-spectral and Breit-Wigner methods 
for different values of the parameter $\mathcal{G}$ and setting $q=0$.
The results of the real and imaginary 
parts of the QN frequency for $\mathcal{G}=1$ are displayed 
in the top panels.
In this figure the numerical results for the fundamental mode ($n=0$)
and $\widetilde{T}^2\ll 1$ were obtained with the Breit-Wigner method,
since the power series method has a poor convergence in this regime.
The opposite occurs as the temperature increases: 
the Breit-Wigner method has a poor convergence while the 
power series method has a good one for high temperatures. 
Such an overlap of results obtained with both methods, in
the intermediate regime of temperatures, is only achieved in
the case of the fundamental mode ($n=0$).
For higher modes ($n\geq 1$), there are some values of the temperature 
for example, $0.04\leq \widetilde{T}^2\leq0.08$,
for which none of these two methods provides solutions.
To fill these empty regions of the spectrum, we use the pseudo-spectral method,
which has a good convergence from the low- to the high-temperature regimes.
For some values in the intermediate regime of temperatures all the 
employed methods work well and it is possible to compare
the numerical results obtained using the three methods.

In the left panels of Fig.~\ref{SpectrumG1G50}, we present the real part of the 
QN frequencies as a function of the temperature. We observe that  
$\widetilde{\omega}_R$ approaches the values displayed
in Table~\ref{Tab:01}, i.e., the spectrum at zero temperature, 
$\widetilde{\omega}^2_{\scriptscriptstyle{R}}
=(4.05\,n+5.17)$ for $\mathcal{G}=1$ and 
$\widetilde{\omega}^2_{\scriptscriptstyle{R}}
=(4.00\,n+5.99)$ for $\mathcal{G}=50$, with $n=0,\,1,\,2,...\,$.  
We also observe a smooth transition between the characteristic
behaviours of low to high temperature regimes.
In particular, $\widetilde{\omega}_R$ scales linearly with $\widetilde{T}$ for high values
of the temperature, i.e., $\widetilde{\omega}_R\propto \widetilde{T}$. 
It is important to point out that, in the low temperature regime, the QN frequencies may be interpreted as quasiparticle states,
since in this case the quality factor is 
greater than one, i.e., $Q=\omega_{R}/(2\omega_{I})\gg 1$. 

In comparison to the real part, the imaginary part of the QN frequencies has a different 
behavior as a function of the temperature: it increases 
monotonically with the temperature, cf. the right panels in
Fig.~\ref{SpectrumG1G50}.
In the limit of zero temperature, the imaginary part
of the frequencies becomes zero, which means that the
lifetime $\tau=1/\omega_{I}$ of the (quasi)particle states 
becomes arbitrarily large.
For intermediate temperatures, $\widetilde{\omega}_{I}$ follows the 
power law $\widetilde{\omega}_I\propto 
\widetilde{T}^\alpha$, where the exponent $\alpha$ can 
be obtained by fitting the numerical results, while for 
high temperatures it grows linearly with the temperature. 

\begin{figure}[!ht]
\centering
\includegraphics[width=8.2cm,angle=0]{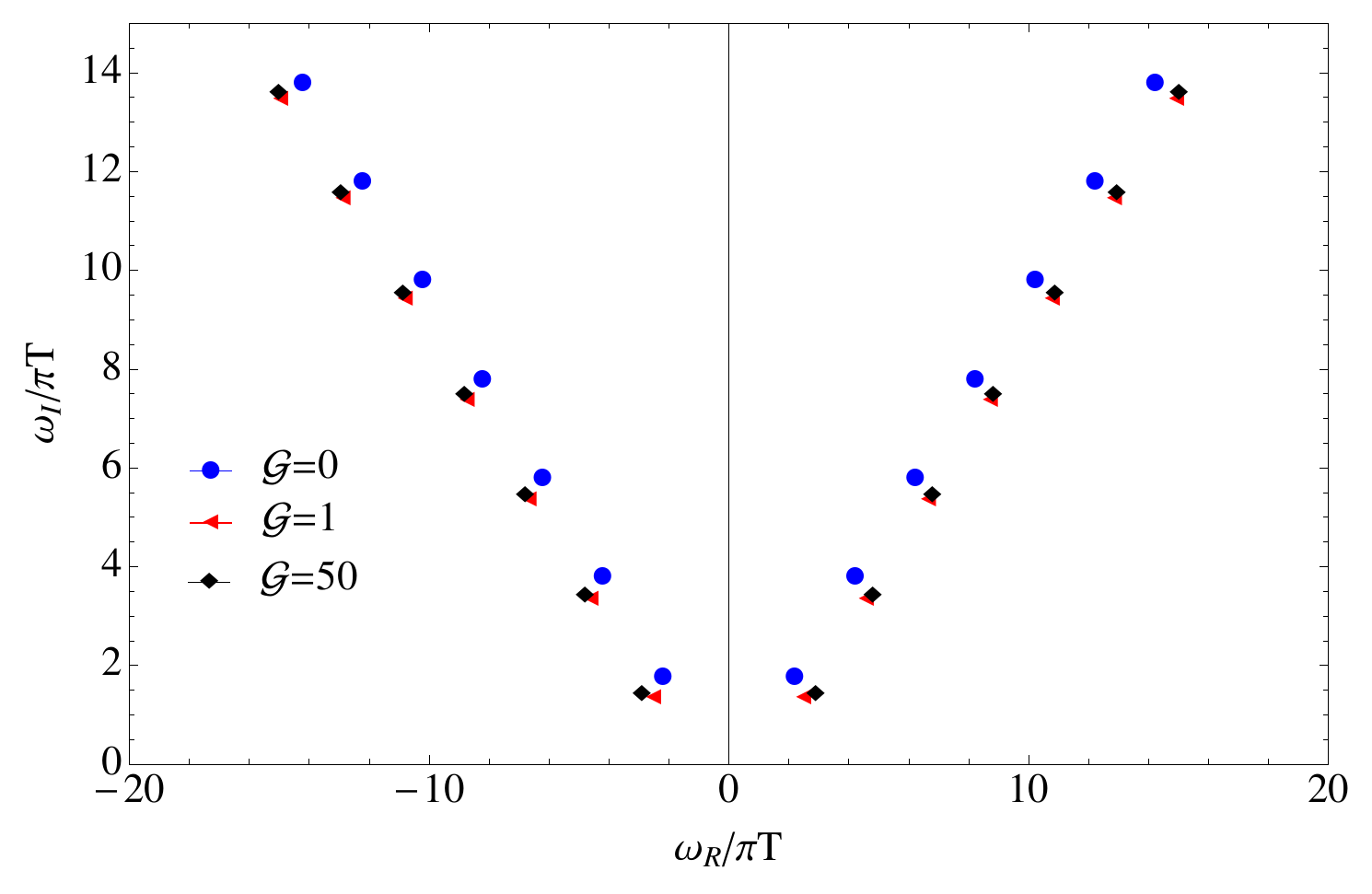} 
\caption{
The the numerical results obtained using the 
pseudo-spectral method 
for selected values of the parameter $\mathcal{G}$.}
\label{PolesStructure}
\end{figure}

Additionally, we plot on the complex plane $\widetilde{\omega}_R \times\widetilde{\omega}_I$ the 
numerical results obtained with the pseudo-spectral method, see Fig.~\ref{PolesStructure}.
It is worth mentioning  that in the case without dilaton, i.e., for, $\mathcal{G}=0$, the results
are identical to those found in Ref.~\cite{Nunez:2003eq}. We may observe in Fig.~\ref{PolesStructure} 
and also in Table \ref{TabQNMsGs} the changes in the structure of poles as a function of the parameter $\mathcal{G}$.

\begin{table*}[ht!]
\centering
\begin{tabular}{ccc|cc|cc}
&\multicolumn{2}{c}{$\mathcal{G}=0$} &
\multicolumn{2}{c}{$\mathcal{G}=1$} &
\multicolumn{2}{c}{$\mathcal{G}=50$}\\ \cline{2-7}
$n$    &
$\omega_{\scriptscriptstyle{R}}/\pi T$ &
$\omega_{\scriptscriptstyle{I}}/\pi T$  &
${\omega}_{\scriptscriptstyle{R}}/\pi T$ &
${\omega}_{\scriptscriptstyle{I}}/\pi T$ &
${\omega}_{\scriptscriptstyle{R}}/\pi T$ &
${\omega}_{\scriptscriptstyle{I}}/\pi T$ \\ 
\hline\hline
0 & $\pm 2.19882$ & 1.75953 & 
    $\pm 2.50343$ & 1.33438 &
    $\pm 2.90009$ & 1.41878 \\
1 & $\pm 4.21190$ & 3.77488 & 
    $\pm 4.59205$ & 3.32799 &
    $\pm 4.80058$ & 3.41043\\
2 & $\pm 6.21554$ & 5.77725 &
    $\pm 6.66143$ & 5.34384 &
    $\pm 6.79148$ & 5.43619\\ 
3 & $\pm 8.21716$ & 7.77808 &
    $\pm 8.72617$ & 7.37164 &
    $\pm 8.82048$ & 7.47670\\ 
4 & $\pm 10.21806$& 9.77847 &
    $\pm 10.79386$& 9.40469 &
    $\pm 10.87444$& 9.51944\\
5 & $\pm 12.21861$& 11.77870&
    $\pm 12.86690$& 11.43744&
    $\pm 12.94509$& 11.55560 \\ 
\hline\hline
 \end{tabular} 
\centering\caption{Numerical results of the first six
quasinormal frequencies obtained using the pseudo-spectral method for
$\widetilde{q}=0$ and selected values of temperature,
by setting $\mathcal{G}=0$, $\mathcal{G}=1$ and $\mathcal{G}=50$.}
\label{TabQNMsGs}
\end{table*}

For completeness we compare the results obtained using
the three numerical methods employed in this work. In 
Table~\ref{TabQNMsCompare} we display some selected values 
of the QN frequencies. We observe that the Breit$-$Wigner and 
the pseudo-spectral methods work well for low temperatures, while 
the power series method has a poor convergence in this regime. Moreover, 
we observe that the Breit$-$Wigner method no longer works well as
the temperature increases, whereas the power series, in contrast,
has a good convergence for high temperatures. It is interesting 
to note that the pseudo-spectral method has a good 
convergence for all the temperatures in 
Table \ref{TabQNMsCompare}. Hence, the pseudo$-$spectral 
method is an alternative to both the Breit$-$Wigner and the power 
series methods. In general, the application of each 
method depends on the subject in which the research is focused.
Finally, we realized that the convergence of these methods 
gets poor as the overtone number increases, $n=2,3,\cdots\,$.

\begin{table*}[ht!]
\centering
\begin{tabular}{ccc|cc|cc}
&\multicolumn{2}{c}{Breit-Wigner} &
\multicolumn{2}{c}{Power Series} &
\multicolumn{2}{c}{Pseudo-spectral}\\ \cline{2-7}
$\widetilde{T}^2$    &
$\widetilde{\omega}_{\scriptscriptstyle{R}}^2$ &
$\widetilde{\omega}_{\scriptscriptstyle{I}}^2$  &
$\widetilde{\omega}_{\scriptscriptstyle{R}}^2$ &
$\widetilde{\omega}_{\scriptscriptstyle{I}}^2$ &
$\widetilde{\omega}_{\scriptscriptstyle{R}}^2$ &
$\widetilde{\omega}_{\scriptscriptstyle{I}}^2$ \\ 
\hline\hline
0.080 & 5.57034 & 1.33179$\times 10^{-3}$
& 5.54109  & 8.11086$\times 10^{-4}$ 
& 5.54446 & 1.23705$\times 10^{-3}$\\
0.075 & 5.60471 & 6.77532$\times 10^{-4}$ 
& 5.59656  & 3.86904$\times 10^{-4}$ 
& 5.59037 & 6.48688$\times 10^{-4}$\\
0.060 & 5.74268 & 3.02717$\times 10^{-5}$ 
& 5.75693 & 3.16718$\times 10^{-5}$ 
& 5.74187 & 3.01802$\times 10^{-5}$\\ 
0.030 & 5.94323 & 4.34800$\times 10^{-15}$ 
& -  & -
& 5.94323 & 4.34811$\times 10^{-15}$\\ 
0.020 & 5.97205 & 1.36357$\times 10^{-19}$ 
& -  & - 
& 5.97205 & 1.08006$\times 10^{-26}$\\
0.010 & 5.98859 & 2.27076$\times 10^{-19}$ 
& -  & - 
& 5.98859 & 1.93069$\times 10^{-27}$ \\ 
\hline\hline
 \end{tabular} 
\centering\caption{Numerical results of the first
quasinormal mode for $\widetilde{q}=0$, $\mathcal{G}=50$ and selected values of temperature.}
\label{TabQNMsCompare}
\end{table*}

\subsection{Comments on the imaginary part of the QN frequencies}

To finish this section we consider some interesting features about the imaginary part 
of the QN frequencies. In particular, we are interested here in the relation
between the equilibration timescale $\tau\propto 1/\omega_I$ and the temperature
of the dual thermal field theory. Let us start with the relation between the imaginary part 
of the frequency and the radius of the event horizon, $r_+$.
In the case of the global Schwarzschild-AdS black hole, it was found in Ref.~\cite{Horowitz:1999jd}
that the imaginary part of the scalar-field QN frequency scales with the horizon area for small 
black holes, i.e., $\omega_I\propto \mathcal{A}$, where $\mathcal{A}$ is the area of the black-hole
event horizon. Moreover, in a five-dimensional spacetime, the area of a spherical region of radius
$r_+$ is such that $\mathcal{A}\propto r_+^{3}$. This behavior was also confirmed with the use of
the Breit-Wigner method in Ref.~\cite{Berti:2009wx} for the scalar-field perturbations
(see Ref.~\cite{Wang:2014eha} for an extension of these results to arbitrary dimensions).
In the case of large black holes, the imaginary part of the QN frequency scales with the horizon radius,
$\omega_{I}\propto r_+$. Hence, we may construct an interpolation function which recovers the
asymptotic behaviors of $\omega_I$. One possibility is
\noindent
\begin{equation}\label{EqImaInter}
{\omega}_I
=\frac{a_1\,r_+^{3}}{b_1+c_1\,r_+^{2}},
\end{equation}
\noindent
where the constants $a_1$, $b_1$ and $c_1$ may be determined by fitting this function to
the numerical results \cite{Horowitz:1999jd}.

The foregoing discussion is valid for Schwarzschild
AdS black holes with spherical geometry.
Now, we address a similar analysis for the planar AdS black hole, which is the case
we are dealing with in this work.
As a matter of fact, our numerical results displayed in Fig.~\ref{SpectrumG1G50} show a linear  
behavior in the regime of large temperatures, $\widetilde{\omega}_I\propto \widetilde{T}$.
This relation, in turn, is no longer valid in the regime of low temperatures. However, we 
know (from Fig.~\ref{SpectrumG1G50}) that the imaginary 
part of the QN frequencies vanishes in the zero temperature limit, so that
the relation must be in the form $\widetilde{\omega}_I\propto \widetilde{T}^{\alpha}$,
where $\alpha$ must be a positive number. Then we construct an interpolation function as  
\noindent
\begin{equation}\label{EqImaInter2}
\widetilde{\omega}_I
=\frac{a\,\widetilde{T}^{\alpha}
+b\,\widetilde{T}^{\alpha+1}}
{c+\,\widetilde{T}^{\alpha-1}},
\end{equation}
\noindent
where the constants may be determined by fitting this
function with our numerical results. Hence, the best 
fit we have found in the interval of temperatures 
$0.01\leq \widetilde T^2\leq 0.5$ is for the values: 
$a=95.0\times 10^{-3}$, $b=1.37$, $c=15.3\times 10^{-4}$ 
and $\alpha=7.60$. This fit was done for the 
case $\mathcal{G}=1$.

\section{Dispersion relations}
\label{Sec:DispersionRelations}

Here we study the momentum dependence of 
the QN frequencies for selected 
values of the temperature. These are the dispersion relations 
$\{\omega_R(q),\omega_I(q)\}$ for the 
scalar field.

\begin{figure*}[!ht]
\centering
\includegraphics[width=7.3cm]{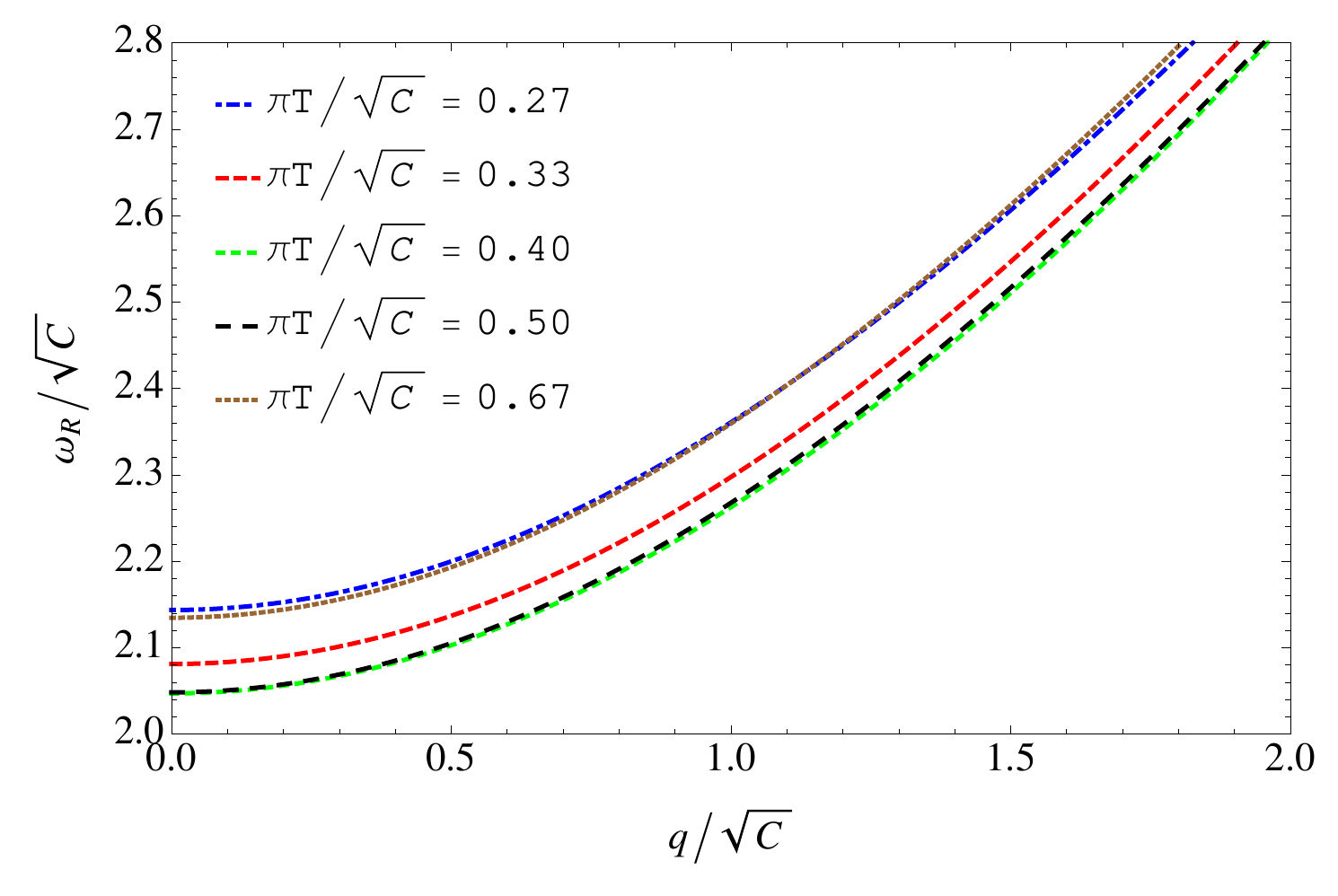}
\hspace{4em}
\includegraphics[width=7.3cm]{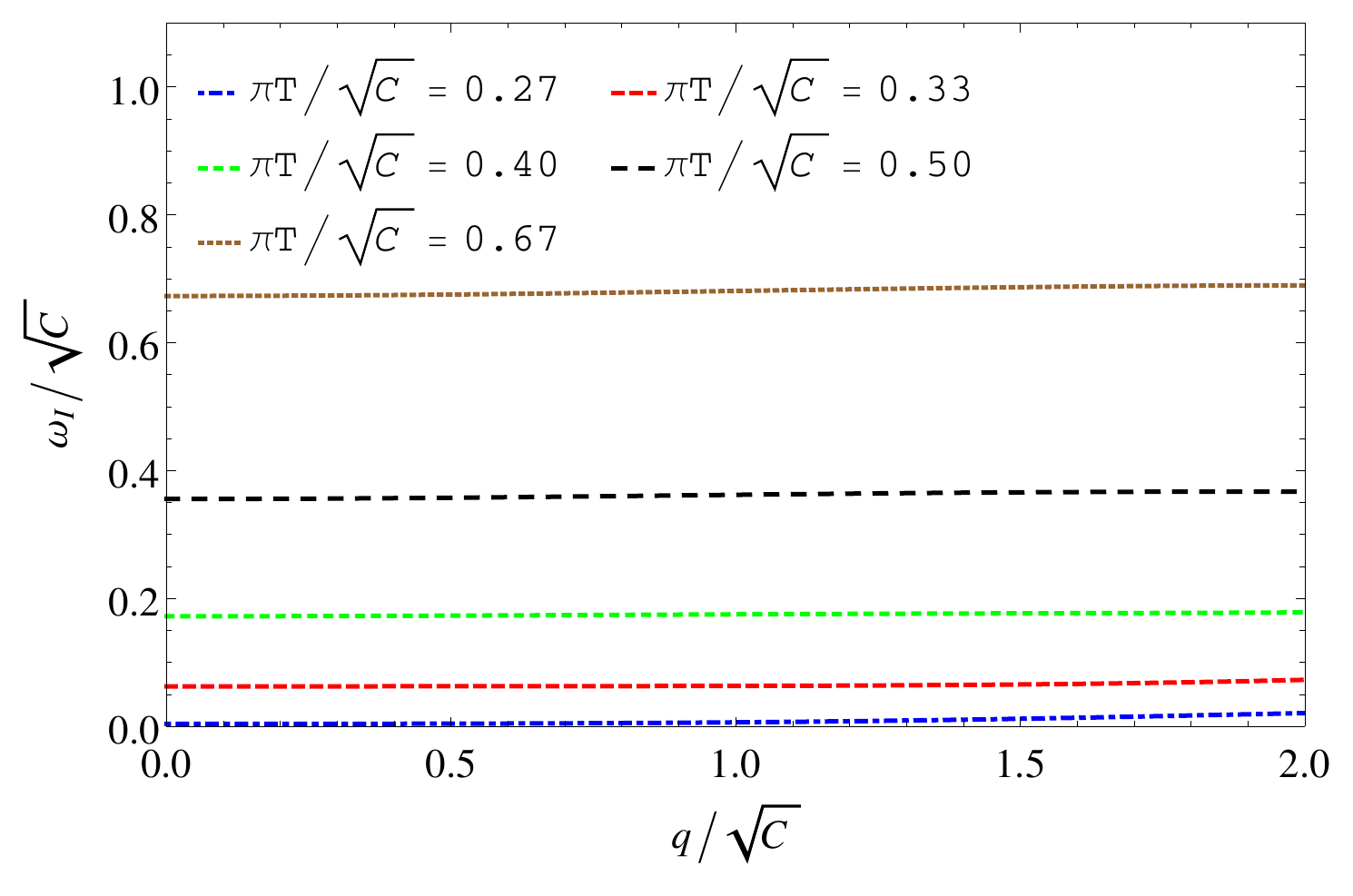}\\
\includegraphics[width=7.3cm]{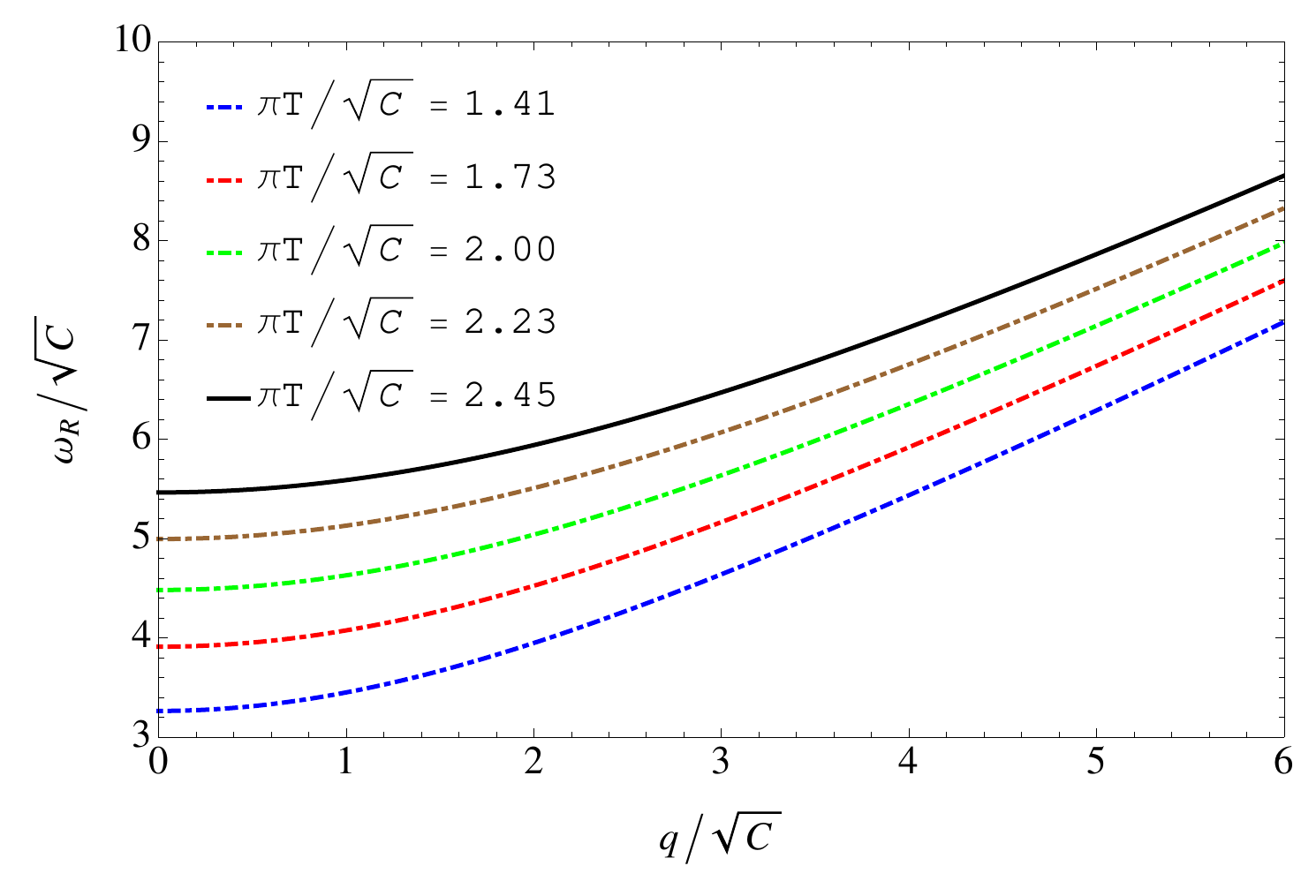}
\hspace{4em}
\includegraphics[width=7.3cm]{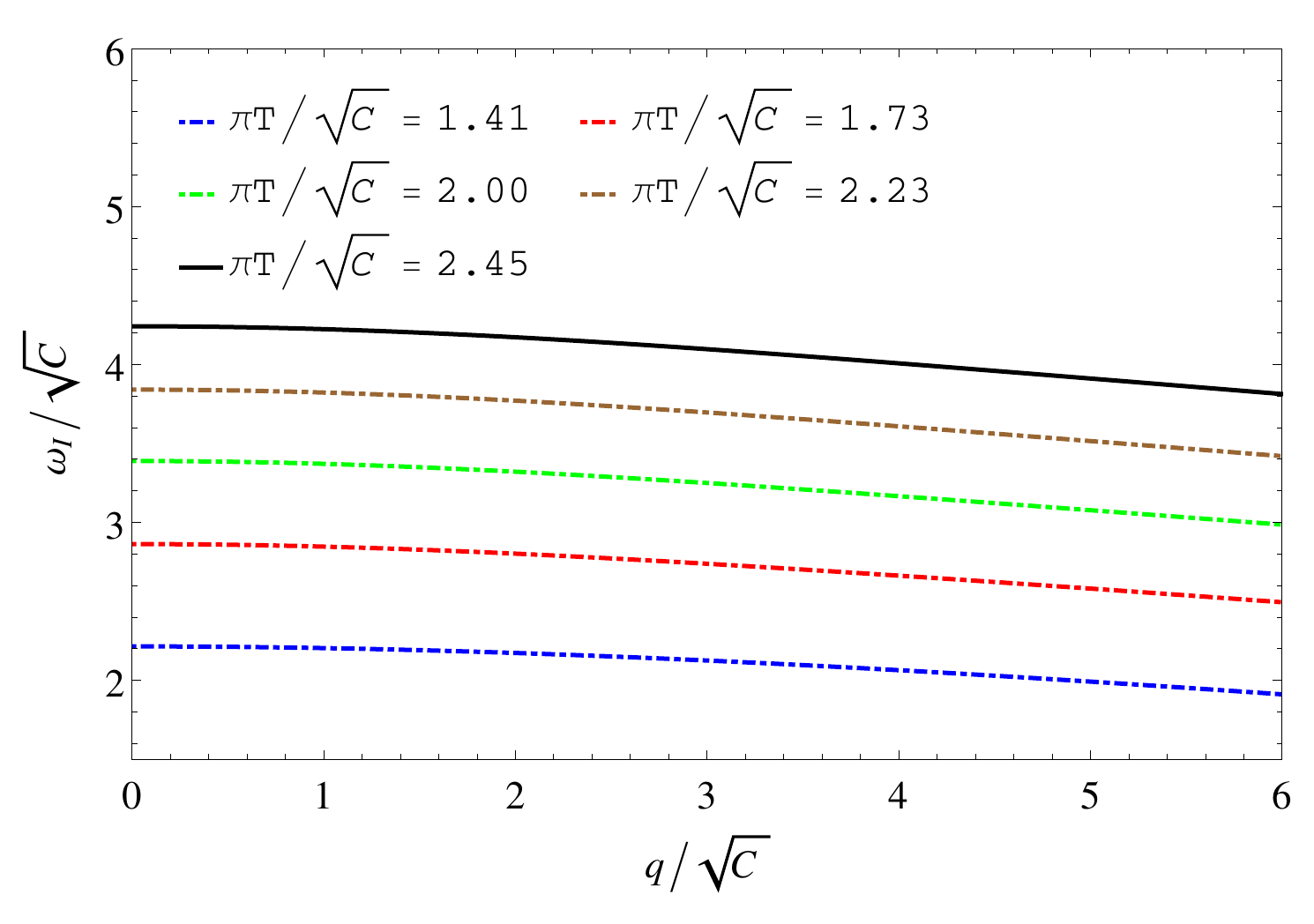}
\caption{Dispersion relations for low temperatures 
(top panels), intermediate and high temperatures (bottom panels) for
$\mathcal{G}=1$.
}\label{DispRelaG1}
\end{figure*}

The dispersion relations for the first five 
QN frequencies, for selected values of temperature and 
setting $\mathcal{G}=1$, are displayed in Fig.~\ref{DispRelaG1}. 
The results for low temperatures are displayed in the top 
panels, while the results for intermediate and high temperatures are displayed in the
bottom panels of the same figure. The obtained frequencies for low temperatures
are consistent with the results presented in Sec.~\ref{Sec:SpectralFunc},
where the real part of the frequency, i.e., the location 
of the peaks of the SPFs, increases with the wavenumber, for the same temperature, cf. 
Fig.~\ref{SpectralFuncLow}. Moreover, the imaginary 
part of the frequency, i.e., the width of the peaks in Fig.~\ref{SpectralFuncLow},
also increases, but very slowly, with the wavenumber. This behaviour holds for low
temperatures, where quasiparticle excitations are present in the quark-gluon plasma.

The results for intermediate and high 
temperatures are displayed in the bottom panels of 
Fig.~\ref{DispRelaG1}. The real part of the 
frequencies grows with the wavenumber, while the imaginary part decreases.
It is worth pointing out that previous studies in the 
literature have calculated QN frequencies of 
massive scalar fields without a dilaton field, and the 
results have a similar behavior for the imaginary 
part of the frequencies,  i.e., $\omega_I$ decreasing with $q$ 
(see for instance Ref.~\cite{Nunez:2003eq}). 
In general, for intermediate and large temperatures, 
this kind of behaviour seems to be an universal property and is shared 
by QNMs in several gravitational theories (see, e.g.,
the reviews of Refs.~\cite{Berti:2009kk,Konoplya:2011qq} and the 
references therein).

\begin{figure*}[ht!]
\centering
\includegraphics[width=7.3cm]{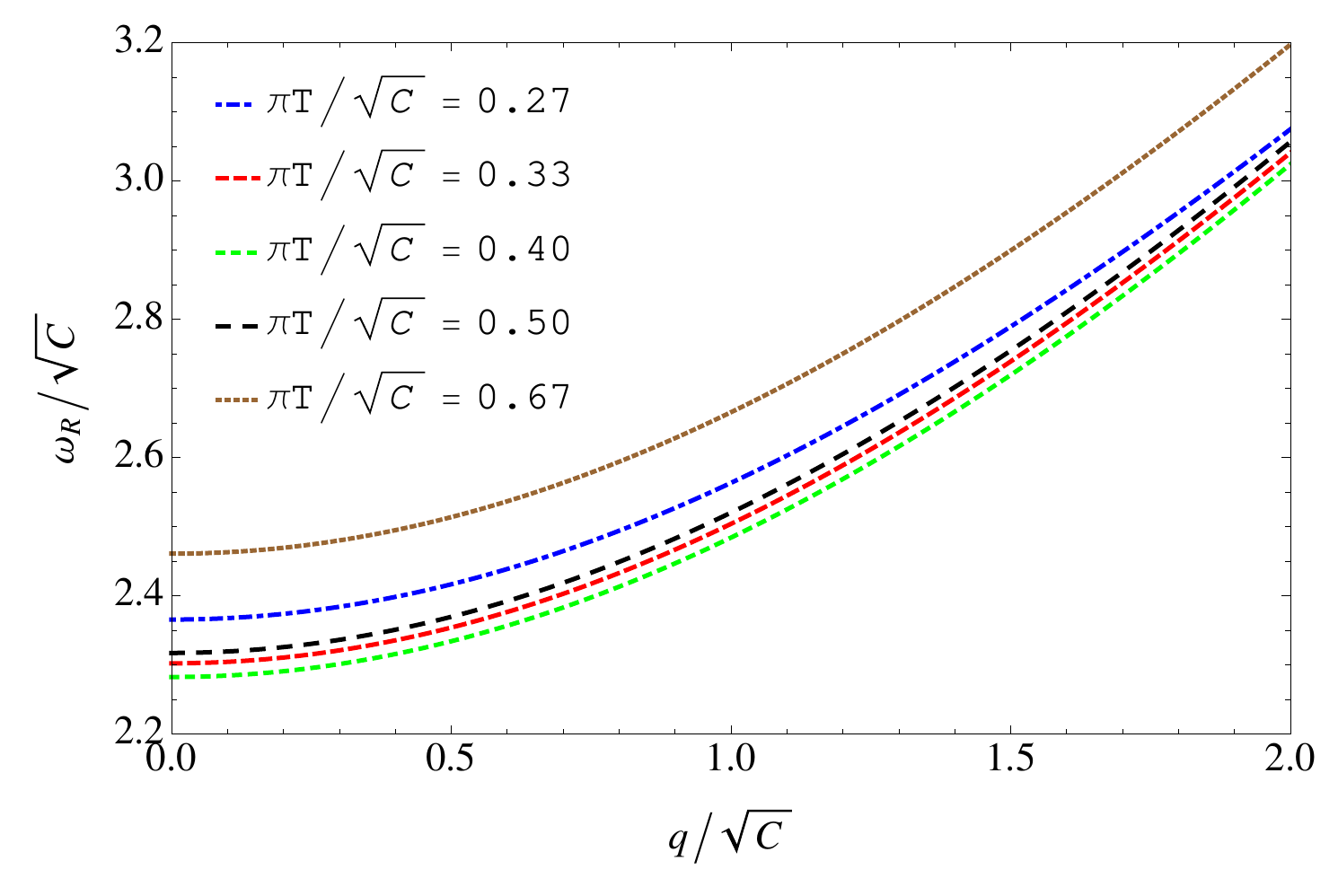}
\hspace{4em}
\includegraphics[width=7.3cm]{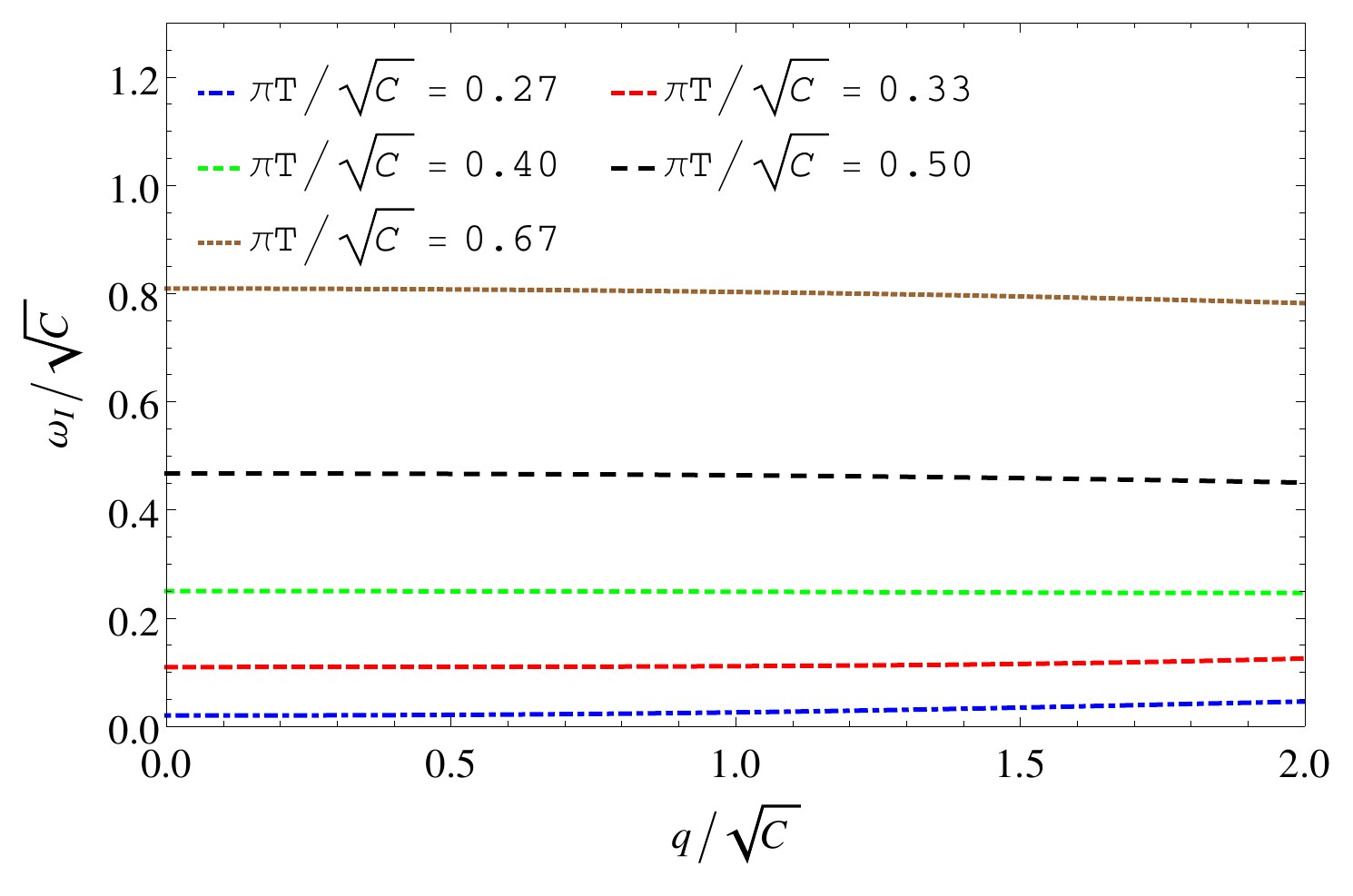}\\
\includegraphics[width=7.3cm]{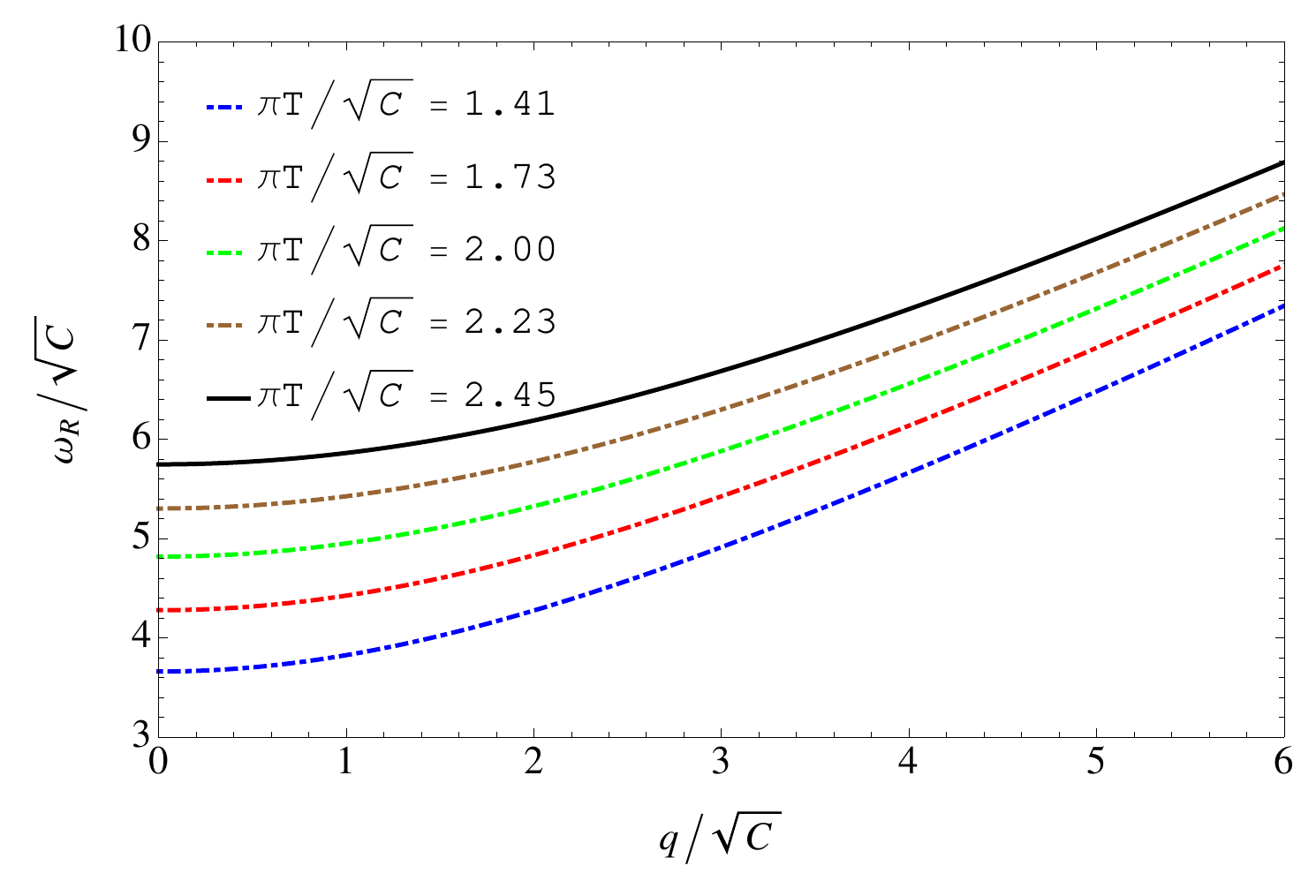}
\hspace{4em}
\includegraphics[width=7.3cm]{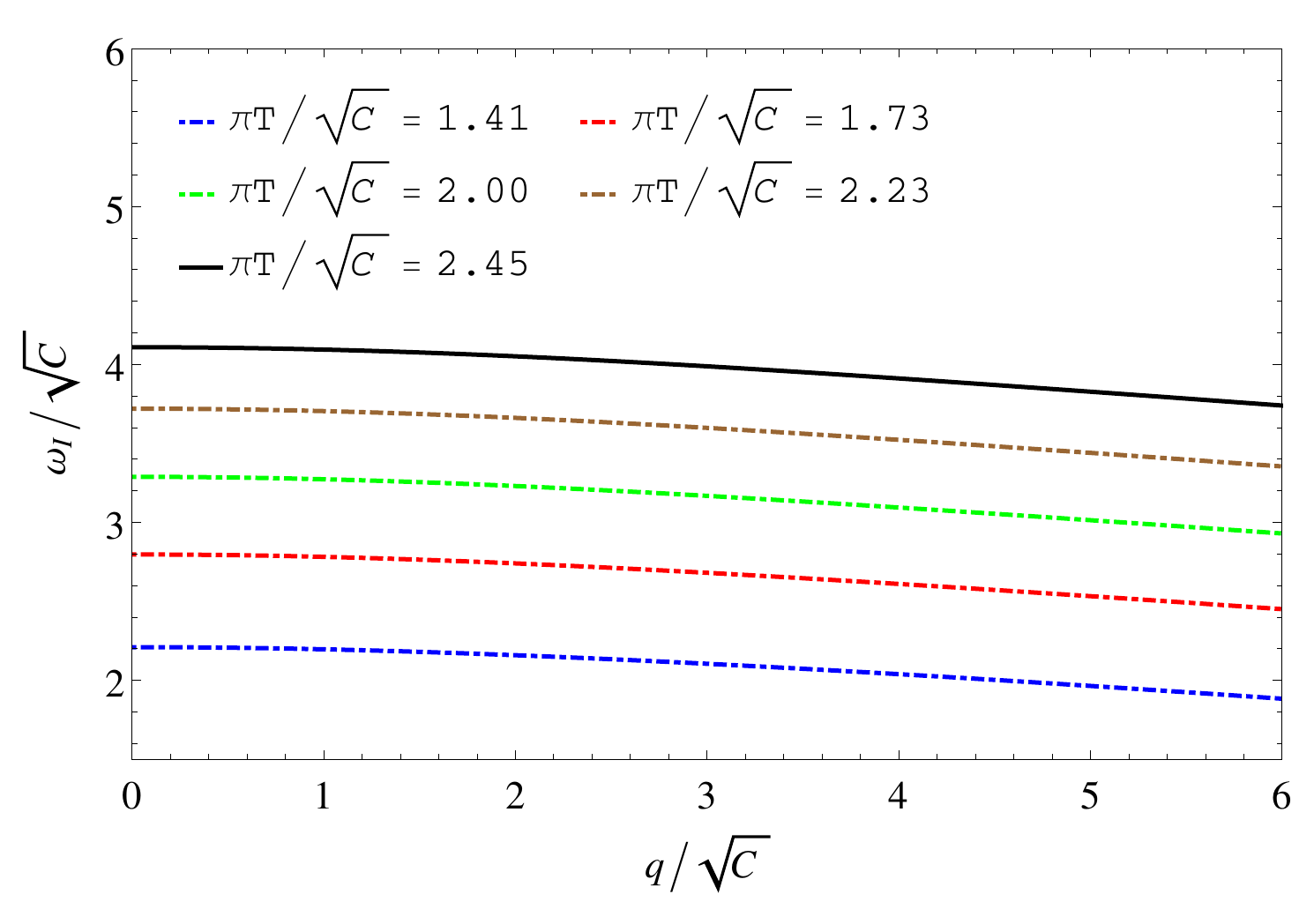}
\caption{Dispersion relations for low temperatures 
(top panels), intermediate and high temperatures (bottom panels) for
$\mathcal{G}=50$.
}\label{DispRelaG50}
\end{figure*}

A second particular case is shown in Fig. \ref{DispRelaG50}, where we plot the QN frequencies
as functions of the normalized wave\-num\-ber $\widetilde{q}$ for $\mathcal{G}=50$.
By comparing the Figs.~\ref{DispRelaG1} and ~\ref{DispRelaG50}, we observe that
the real part of the frequencies calculated for $\mathcal{G}=50$ 
is larger than the one calculated for $\mathcal{G}=1$ 
(for the same low temperature). The same is true for the 
imaginary part of the frequencies,
$\omega_{I}(\mathcal{G}=50)>\omega_{I}(\mathcal{G}=1)$, which means
that the thermalization timescales $\tau=1/\omega_{I}$
become smaller as the value of the parameter $\mathcal{G}$ increases
(see also Table~\ref{TabDRLowT}). Observing carefully the top right panel 
of Fig.~\ref{DispRelaG50} we realized that for 
$\widetilde{T}\geq 0.40$ the imaginary part of the frequency decreases 
slightly, and this is different from the corresponding result 
for $\mathcal{G}=1$. On the other hand, for higher values of temperature, 
the behavior of the imaginary part of the QN frequencies is the opposite, 
$\omega_{I}(\mathcal{G}=50)<
\omega_{I}(\mathcal{G}=1)$, which means that the 
thermalization timescales are shorter for small 
$\mathcal{G}$ (see also Table~\ref{TabDRHighT}).

To explicitly verify the dependence of the QN frequencies on the temperature
we show in Tables \ref{TabDRLowT} and 
\ref{TabDRHighT} some selected numerical results for a fixed wavenumber value $\widetilde{q}=1$,
and the same two values of parameter $\mathcal{G}$ of Figs.~\ref{DispRelaG1} and \ref{DispRelaG50}.
The dependence on the temperature is seen by comparing the two tables, while the dependence
on the values of the parameter $\mathcal{G}$ is seen within each table.

\begin{table}
\centering
\begin{tabular}{ccc|cc}
&\multicolumn{2}{c}{$\mathcal{G}=1$} &
\multicolumn{2}{c}{$\mathcal{G}=50$} \\ \cline{2-5}
$\widetilde{T}$    &
$\widetilde{\omega}_{\scriptscriptstyle{R}}$ &
$\widetilde{\omega}_{\scriptscriptstyle{I}}$  &
$\widetilde{\omega}_{\scriptscriptstyle{R}}$
& $\widetilde{\omega}_{\scriptscriptstyle{I}}$ \\ 
\hline\hline
0.27 & 2.36056 & 0.00611 & 2.56318  & 0.02563 \\
0.33 & 2.29729 & 0.06309 & 2.50387  & 0.11104 \\
0.40 & 2.26243 & 0.17505 & 2.48310  & 0.24863 \\
0.50 & 2.26746 & 0.36195 & 2.52008  & 0.46387 \\ 
0.67 & 2.35912 & 0.68080 & 2.66534  & 0.80273 \\ 
\hline\hline
 \end{tabular} 
\centering\caption{Numerical results of the first quasinormal mode for
$\widetilde{q}=1$ and low temperatures.}
\label{TabDRLowT}
\end{table}

\begin{table}[ht]
\centering
\begin{tabular}{ccc|cc}
&\multicolumn{2}{c}{$\mathcal{G}=1$} &
\multicolumn{2}{c}{$\mathcal{G}=50$} \\ \cline{2-5}
$\widetilde{T}$    &
$\widetilde{\omega}_{\scriptscriptstyle{R}}$ &
$\widetilde{\omega}_{\scriptscriptstyle{I}}$  &
$\widetilde{\omega}_{\scriptscriptstyle{R}}$
& $\widetilde{\omega}_{\scriptscriptstyle{I}}$ \\ 
\hline\hline
1.41 & 3.44958  & 2.20297 & 3.82511 & 2.19533 \\
1.73 & 4.07448  & 2.84548 & 4.42452 & 2.78078 \\
2.00 & 4.62865  & 3.36980 & 4.95030 & 3.27206 \\
2.23 & 5.12904  & 3.82111 & 5.42478 & 3.70372 \\ 
2.45 & 5.58760  & 4.22248 & 5.86061 & 4.09311 \\ 
\hline\hline
 \end{tabular} 
\centering\caption{Numerical results of the first quasinormal mode for
$\widetilde{q}=1$ and high temperatures.}
\label{TabDRHighT}
\end{table}

\section{Final remarks and conclusion}
\label{Sec:conclusion}

In this work we have considred the effects of the gluon
condensate in the spectrum and melting of 
scalar mesons in holographic QCD. 
Additionally, in the gravitational background, we 
determined the spectrum of QNMs and explore the dependence of 
such a spectrum on the parameter 
associated with the gluon condensate.
The effects of the gluon condensate were introduced by considering a 
quartic dilaton in the UV and quadratic in the IR (in order to 
guarantee confinement and linear Regge trajectories). The 
results obtained show that the spectrum
at zero temperature is sensitive to the value of the
energy scale $G$ associated with the condensate.
For large values of this parameter we 
obtain the same spectrum as obtained in
the original soft-wall model applied to the case of 
scalar mesons.
On the other hand, when this parameter is zero 
we recover the problem of an AdS spacetime with 
constant dilaton field. In this case the conformal 
symmetry is restored and the spectrum becomes continuum. 

The results are more interesting when the gravitational background contains a black hole, since
the presence of the black hole implies in a dual field theory at finite temperature.
In this case the conformal symmetry is broken by the dilaton and by the temperature.
Differently from the zero temperature case, even in the case $G=0$ 
and the results show a discrete spectrum
because the potential has a well due to finite temperature. Our results also show that
the potential well depends directly on the 
dimensionless parameters $\mathcal{G}$ and $\widetilde{T}$. 
These parameters deform the potential in such a way
that there are more trapped quasiparticle states at low temperatures, where the potential 
has wells (cf. Fig.~\ref{PotentialLow}). These results are
supported nicely by the spectral functions (cf.~\ref{SpectralFuncLow}). 
In the low-temperature regime, we observe the presence of peaks in the 
spectral functions, and such peaks disappear when the temperature 
increases, characterizing the melting of the scalar mesons. We 
also observe that higher excited quasiparticle states melt faster than low excited states
for the same temperature. 
These effects are enhanced when spatial momentum is taken into account.

We complement this work by calculating the spectrum of black-hole
quasinormal modes associated to the scalar-field perturbations. The
numerical results show a finite real part of the frequencies in the
limit of zero temperature, while the imaginary part becomes zero
(cf. Fig.~\ref{SpectrumG1G50}). In this limit the QNMs 
become normal modes. As an additional analysis we have
used three numerical techniques to calculate the 
spectrum, and from a comparison, we realized that they work well 
in a determinate regime of temperatures and can be used for specific purposes.
At the end we present the numerical results and a discussion on the 
dispersion relations. In all the results obtained in this 
work we observe the relevance of the parameter 
$\mathcal{G}$, and how important is to take it into account 
when studying the melting of scalar mesons in a holographic 
model for QCD.

\begin{acknowledgements}
We thank Alfonso Ballon-Bayona for stimulating discussions 
along the development of this work, we also 
thank Carlisson Miller and Saulo Diles for reading and comments 
on the manuscript.
L. A. H. M. thanks financial support from Funda\c{c}\~ao de
Amparo \`a Pesquisa do Estado de S\~ao Paulo (FAPESP, Brazil), Grant No. 
2013/17642-5, and from Coordena\c{c}\~ao de Aperfei\c{c}oamento do Pessoal de N\'ivel 
Superior - Programa Nacional de P\'os-Doutorado 
(PNPD\-/\-CAPES, Brazil).
V. T. Z. thanks financial support from Conselho Nacional de Desenvolvimento Cient\'\i fico
e Tecnol\'ogico (CNPq, Brazil), Grant No.~308346/2015-7, and from Coordena\c{c}\~ao de
Aperfei\c{c}oamento do Pessoal de N\'\i vel Superior (CAPES, Brazil),
Grant No. 8\-8\-8\-8\-1.0\-6\-4\-9\-9\-9/2014-01.
\end{acknowledgements}

\end{document}